\documentclass[aoas,preprint]{imsart}

\RequirePackage[OT1]{fontenc}
\RequirePackage{amsthm,amsmath}
\RequirePackage{natbib}
\RequirePackage[colorlinks,citecolor=blue,urlcolor=blue]{hyperref}

\startlocaldefs
\numberwithin{equation}{section}
\theoremstyle{plain}

\endlocaldefs

\usepackage{amsmath}
\usepackage{bbm}
\usepackage{multirow}
\usepackage{graphicx}
\DeclareGraphicsExtensions{.pdf,.jpg,.png}
\usepackage{subfigure}

\usepackage{amsfonts}
\usepackage{algorithm,algorithmicx,algpseudocode}

\begin{document}

\begin{frontmatter}
\title{Gene network reconstruction using global-local shrinkage priors\thanksref{T1}}
\runtitle{Network inference using global-local shrinkage priors}

\thankstext{T1}{This work was supported by the Center for Medical Systems Biology (CMSB), established by the Netherlands Genomics Initiative/Netherlands Organization for Scientific Research (NGI/NWO), and the European Union Grant “EpiRadBio”, nr. FP7-269553.}

\begin{aug}
\author{\snm{Gwena\"el G.R. Leday}\thanksref{m1}\ead[label=e1]{gwenael.leday@mrc-bsu.cam.ac.uk}},
\author{\snm{Mathisca C.M. de Gunst}\thanksref{m2}\ead[label=e2]{degunst@cs.vu.nl}},
\author{\snm{Gino B. Kpogbezan}\thanksref{m3}\ead[label=e4]{g.b.kpogbezan@math.leidenuniv.nl}},
\author{\snm{Aad W. Van der Vaart}\thanksref{m3}\ead[label=e3]{avdvaart@math.leidenuniv.nl}},
\author{\snm{Wessel N. Van Wieringen}\thanksref{m2,m4}\ead[label=e5]{w.vanwieringen@vumc.nl}}
\and
\author{\snm{Mark A. Van de Wiel}\thanksref{m2,m4}\ead[label=e6]{mark.vdwiel@vumc.nl}}

%\thankstext{t1}{Some comment}
%\thankstext{t2}{First supporter of the project}
%\thankstext{t3}{Second supporter of the project}
\runauthor{Leday et al.}

\affiliation{MRC Biostatistics Unit\thanksmark{m1}, Vrije Universiteit Amsterdam\thanksmark{m2}, \\Leiden University\thanksmark{m3} and VU University Medical Center\thanksmark{m4}}

\address{MRC Biostatistics Unit\\Cambridge Institute of Public Health\\Forvie Site\\Robinson Way\\Cambridge Biomedical Campus\\Cambridge CB2 0SR\\United Kingdom\\\printead{e1}}

\address{Department of Mathematics\\Vrije Universiteit Amsterdam\\De Boelelaan 1081\\1081 HV Amsterdam\\The Netherlands\\ \printead{e2}\\ \printead{e5}\\\printead{e6}}

\address{Mathematical Institute\\Faculty of Science\\Leiden University\\P.O. Box 9512\\2300 RA Leiden\\The Netherlands\\ \printead{e3}\\ \printead{e4}}

\address{Department of Epidemiology and Biostatistics\\ VU University Medical Center\\ PO Box 7057\\ 1007 MB Amsterdam\\ The Netherlands\\ \printead{e5}\\ \printead{e6}}

\end{aug}

\begin{abstract}
Reconstructing a gene network from high-throughput molecular data is often a challenging task, as the number of parameters to estimate easily is much larger than the sample size. A conventional remedy is to regularize or penalize the model likelihood. In network models, this is often done \emph{locally} in the neighbourhood of each node or gene. However, estimation of the many regularization parameters is often difficult and can result in large statistical uncertainties. In this paper we propose to combine local regularization with \emph{global} shrinkage of the regularization parameters to borrow strength between genes and improve inference. We employ a simple Bayesian model with non-sparse, conjugate priors to facilitate the use of fast variational approximations to posteriors. We discuss empirical Bayes estimation of hyper-parameters of the priors, and propose a novel approach to rank-based posterior thresholding. Using extensive model- and data-based simulations, we demonstrate that the proposed inference strategy outperforms popular (sparse) methods, yields more stable edges, and is more reproducible.
\end{abstract}

%\begin{keyword}[class=MSC]
%\kwd[Primary ]{60K35}
%\kwd{62P10}
%\kwd[; secondary ]{60K35}
%\end{keyword}


\end{frontmatter}

%%%%%%%%%%%%%%%%%%%%%%%%%%%%%%%%%%%%%%%%%%%%%%%%%%%%%%
%%%%%%%%%%%%%%%%%%%%%%%%%%%%%%%%%%%%%%%%%%%%%%%%%%%%%%
%%%	section: INTRODUCTION
%%%%%%%%%%%%%%%%%%%%%%%%%%%%%%%%%%%%%%%%%%%%%%%%%%%%%%
%%%%%%%%%%%%%%%%%%%%%%%%%%%%%%%%%%%%%%%%%%%%%%%%%%%%%%

\section{Introduction}
Gaussian Graphical Models (GGMs) are a popular tool in genomics to describe functional dependencies between biological units of interest, such as genes or proteins. These models provide means to apprehend the complexity of molecular processes using high-throughput experimental data, and shed light on key regulatory genes or proteins that may be interesting for further follow-up studies. Among the many approaches that have been advanced, simultaneous-equation models (SEMs), which express each gene or protein expression profile as a function of other ones, have been found particularly valuable owing to their flexibility and simplicity. Notably, SEMs facilitate \emph{local} regularization,  where for each gene the set of parameters that model its dependence on the other genes is penalized separately and possibly to a different amount. However this comes at the price of having many regularization parameters, which may be difficult to tune. Motivated by works in the field of differential expression analysis, in this paper we combine local regularization with \emph{global} shrinkage of the regularizing parameters to stabilize and improve estimation. Adopting a Bayesian approach, we demonstrate, using extensive model- and data-based simulations, that such global shrinkage may substantially improve statistical inference.

GGMs characterize the dependence structure between molecular variables using partial correlations. It is well known that two coordinates $Y_i$ and $Y_j$ of a multivariate normal random vector $Y=(Y_1,\ldots,Y_p)^T$ are conditionally independent given the set of all other coordinates  if and only if the partial correlation corr$(Y_i,Y_j|Y_{\mathcal{J}\setminus\{i,j\}})$ is zero, where $\mathcal{J}=\{1,\ldots,p\}$. Furthermore, if $Y\sim\mathcal{N}_p(0, \Omega^{-1})$ with positive-definite \emph{precision matrix}
$\Omega=(\omega_{ij})$, then these partial correlations can be expressed as corr$(Y_i,Y_j|Y_{\mathcal{J}\setminus\{i,j\}})=-\omega_{ij}/\sqrt{\omega_{ii}\omega_{jj}}$, for $i\neq j$. Thus the conditional dependence structure is fully coded in the precision matrix, and a network structure may be defined by discriminating the zero and non-zero entries of the precision matrix. It is convenient to represent this structure by an undirected graph $\mathcal{G}=\{\mathcal{J}, \mathcal{E}\}$, with the nodes $\mathcal{J}$ corresponding to the variables, and the edge set $\mathcal{E}$ consisting of all $\{i,j\}$ such that $\omega_{ij}\neq 0$. 

Most modern inference techniques for GGMs focus on estimating $\Omega$ or this underlying graph. For brevity we only discuss the most popular methods, which will also be used as benchmarks in our simulations.

Penalized likelihood estimation amounts to maximizing 
$\ell (\Omega) = \log|\Omega| - tr(S\Omega) - \lambda J(\Omega)$, where $S$ is the sample covariance estimate, $J$ a penalty function, and $\lambda$ a scalar tuning parameter. The penalty $J$ may serve two purposes: (1) to ensure identifiability and improve the quality of estimation; (2) to discriminate zero from non-zero entries in $\Omega$. The $\ell_{1}$-norm  (or versions thereof) is a popular choice \citep{friedman2008}, because it simultaneously achieves (1) and (2). Alternatively, a ridge-type penalty \citep{vanWieringen2014,warton2008,ledoit2004} may be used in combination with a thresholding procedure \citep{schafer2005}.
Appropriate tuning of the penalty through the parameter $\lambda$ is crucial for good performance. Various solutions, usually based on resampling or cross-validation, have been proposed \citep{gao2012,lian2011,meinshausen2010,foygel2010,giraud2008,yuan2007}.

Simultaneous-equation modelling estimates $\Omega$ by regressing each molecular variable $Y_j$ against all others. The coefficients $\beta_{j,k}$ in the equations
\begin{equation}
	\label{NR}
		Y_j = \sum_{k\in \mathcal{J}\setminus j}{Y_{k} \beta_{j,k}} + \epsilon_j,\quad j\in \mathcal{J},
\end{equation}
where $\epsilon_j \sim \mathcal{N}(0,\sigma_j^2)$ is independent of $(Y_k: k\not=j)$, can be shown to be given by $\beta_{j,k}=-\omega_{jj}^{-1}\omega_{jk}$. Also $\sigma_j^{2}=\omega_{jj}^{-1}$. Consequently, identifying the nonzero entries of $\Omega$  can be recast as a variable selection problem in $p$ Gaussian regression models. This approach to graphical modeling was popularized by \citet{meinshausen2006}. They dealt with high-dimensionality by adding an $\ell_{1}$-penalty to each regression problem, but other penalties are also used \citep{kramer2009}. Because the model \eqref{NR} misses the symmetry
$\omega_{ij}=\omega_{ji}$ in $\Omega$, estimation may lack efficiency. This may be overcome by working directly on partial correlations, as shown by \citet{peng2009}. Alternatively,
\citet{meinshausen2006} proposed a \textit{post-symmetrization} step with an `AND' rule: edge $(i,j)\in\mathcal{E}$ if $\beta_{i,j}\neq 0$ and $\beta_{j,i}\neq 0$. Despite the symmetry issue, network reconstruction using \eqref{NR} performs well and is widely used in practice.

Simultaneous-equation models are quite flexible. Experimental or biological covariates can easily be accounted for in the regression, and extensions to non-Gaussian data were suggested by 
\citet{chen2015,allen2013,yang2012,ravikumar2010}. Also SEMs arise naturally from the differential equations of a general dynamical system model of gene regulation \citep{oates2012}.

In this paper we develop a Bayesian approach to Gaussian graphical modeling using SEMs. Our contribution is three-fold: (1) we employ \eqref{NR} in combination with (non-sparse) priors that induce both \emph{local} and \emph{global} shrinkage and provide evidence that global shrinkage may substantially improve inference; (2) we present a new approach to posterior thresholding using a concept similar to the local false discovery rate \citep{efron2010} and show that non-sparse priors coupled with a posteriori edge selection are a simple and attractive alternative to sparse priors; and (3) we provide a computationally attractive software tool called \texttt{ShrinkNet} (available at \url{http://github.com/gleday/ShrinkNet}), which is based on a coherent and complete estimation procedure that does not rely on resampling or cross-validation schemes to tune parameter(s).

The paper is organized as follows. Section~\ref{meth} presents the Bayesian SEM, the variational approximation to posteriors and a novel posterior thresholding procedure to reconstruct the network. In this section we also describe estimation of the global shrinkage prior and discuss the important role of the proposed empirical Bayes procedure, along with its connection to existing literature. In Sections~\ref{modelSim} and~\ref{dataSim} we compare the performance of the new method with state-of-the-art sparse and non-sparse approaches, using both model- and data-based simulations. Notably in Section~\ref{dataSim} we employ two mRNA expression data sets from The Cancer Genome Atlas (TCGA) and a random-splitting strategy 
to compare the reproducibility and stability of the various methods.

%%%%%%%%%%%%%%%%%%%%%%%%%%%%%%%%%%%%%%%%%%%%%%%%%%%%%%
%%%%%%%%%%%%%%%%%%%%%%%%%%%%%%%%%%%%%%%%%%%%%%%%%%%%%%
%%%	section: METHODS
%%%%%%%%%%%%%%%%%%%%%%%%%%%%%%%%%%%%%%%%%%%%%%%%%%%%%%
%%%%%%%%%%%%%%%%%%%%%%%%%%%%%%%%%%%%%%%%%%%%%%%%%%%%%%

\section{Methods}
\label{meth}
In this section we introduce the Bayesian SEM with global and local shrinkage priors along with a variational approximation of the resulting posterior distribution(s). Next we present empirical Bayes estimation of prior hyper-parameters. We conclude with a selection procedure for inferring the edge set $\mathcal{E}$.

%%%%%%%%%%%%%%%%%%%%%%%%%%%%%%%%%%
%%% BAYESIAN SEM
%%%%%%%%%%%%%%%%%%%%%%%%%%%%%%%%%%

\subsection{The Bayesian SEM}
\label{meth:model}
Consider mRNA expression data on $p$ genes from $n$ sample tissues.
Denote by $\mathbf{y}_j$ the $n\times 1$ vector of mRNA expression ($\log_2$) values for  gene $j\in \mathcal{J}=\{1,\ldots,p\}$. The Bayesian SEM is defined by equation \eqref{NR}
together with a hierarchical specification of prior distributions:
\begin{equation}
	\label{BSEM}
	\begin{split}
	\mathbf{y}_j& = \sum_{k\in \mathcal{J}\setminus j}{\mathbf{y}_k \beta_{jk}}  + \boldsymbol{\epsilon}_j, \qquad j=1,\ldots,p\\
	\boldsymbol{\epsilon}_j &\sim \mathcal{N}_n(0,\sigma_j^2 \mathbf{I}_n),\\
	\beta_{jk}&\sim \mathcal{N}(0, \sigma_j^2\tau_j^2),\\
	\tau_j^{-2}&\sim \mathcal{G}(a, b),\\
	\sigma_j^{-2}&\sim \mathcal{G}(c, d).
	\end{split}
\end{equation}
Here (as usual) every line is understood to be conditional on the lines below it and variables within a line are assumed independent, as are variables referring to different genes $j$. Furthermore, $\mathcal{G}\left(s, r\right)$ denotes a gamma distribution with shape and rate parameters $s$ and $r$, and $\mathbf{I}_n$ is the $n\times n$ identity matrix. Throughout the paper the hyper-parameters $c$ and $d$ are fixed to small values, e.g.\ $0.001$, in contrast to $a$ and $b$, which we will estimate (see Section \ref{meth:empBayes}). Although $c$ and $d$ could also be estimated, we prefer a non-informative prior for the parameters $\sigma_j$, as there seems no reason to connect the error variances across the equations.

The regression parameters $\beta_{jk}$ are endowed with gene-specific, Gaussian priors for \emph{local} shrinkage. A small value of the prior variance $\tau_j^2$ encourages the posterior distributions of the $\beta_{jk}$ (including their expectations $\mathbb{E}(\beta_{jk}\vert\mathbf{y}_j)$) to be shrunken towards zero. The stabilizing effect of this ridge-type shrinkage has been observed to be useful for ranking regression parameters as a first step in variable selection \citep{bondell2012}. In Section~\ref{meth:selection} we show how similarly the marginal posterior distributions of the $\beta_{jk}$ can be used for rank-based edge selection in a GGM. The prior variances of the $\beta_{jk}$ are also defined proportional to the error variances $\sigma_j^2$ to bring the variances $\tau_j^2$, and the induced shrinkage, on a comparable scale \citep{park2008}.

The equations for different genes $j$ are connected through the gamma priors placed on the precisions $\tau_{j}^{-2}$ and the error variances $\sigma_j^2$, for $j\in\mathcal{J}$. The prior on the error variances has no structural role, and, as mentioned, we prefer a fixed non-informative prior. In contrast, the $\mathcal{G}(a, b)$-prior on the precisions $\tau_j^{-2}$ induces \emph{global} shrinkage by borrowing strength across the regression equations. The \emph{exchangeability} of the precisions expressed through this
prior acknowledges the fact that the equations for the different genes are similar in a broad sense, which is plausible given that they share many common elements. When informative (i.e.\ small or moderate value of $a/b^2$), this prior shrinks the posterior distributions of $\tau_{j}^{-2}$ towards the prior mean $a/b$, which 
stabilizes estimation. This type of shrinkage is different from the shrinkage of the regression coefficients $\beta_{jk}$, which through their centered priors are always shrunken to zero. Of course, the ``informed'' shrinkage of the precisions $\tau_j^{-2}$ will be beneficial only if the hyper parameters $a$ and $b$ are chosen appropriately. We propose to set their values based on 
the data, using an empirical Bayes approach, discussed in Section~\ref{meth:empBayes}.

The conjugacy of the Gaussian and gamma priors in model \eqref{BSEM} confers the method a computational advantage over complex sparse priors. Fast approximations to the posteriors are readily available \citep{rajagopalan1983,rue2009,omerod2010}, whereas sparse, non-conjugate priors often require MCMC. The Gaussian priors allow to reparameterize the problem employing an SVD decomposition of the design matrix \citep{west2003}, and back-transform the posteriors to the original space (at least in our setting with approximately Gaussian posteriors; see Section~\ref{meth:VB}), which is computationally advantageous when $p>n$.

A disadvantage of these priors is that they do not have an intrinsic variable selection property, whence the posterior does not automatically recover the graph structure. We solve this by a separate procedure for variable selection, which essentially consists of thresholding the scaled posterior means of the regression coefficients $\beta_{jk}$. In Section~\ref{meth:selection} we present an approach based on Bayes factors and a local false discovery rate.

%%%%%%%%%%%%%%%%%%%%%%%%%%%%%%%%%%
%%% VARIATIONAL APPROXIMATION
%%%%%%%%%%%%%%%%%%%%%%%%%%%%%%%%%%
\subsection{Variational approximation to posteriors}
\label{meth:VB}
Because intractable integrals make it difficult to obtain the exact posterior distribution of the parameters, we use a variational approximation. Variational inference is a fast deterministic
alternative to MCMC methods, and consists of computing a best approximation to the posterior distribution from a prescribed family of distributions. In our situation it provides an analytic expression for a lower bound on the log-marginal likelihood, which is useful
for monitoring convergence of the algorithm and to assess model fit (Section \ref{meth:empBayes}).

For given hyper-parameters $(a,b)$ and with the variables $\mathbf{y}_k$ in the right side of \eqref{BSEM} considered fixed covariates, the prior and posterior distributions factorize (i.e.\ are independent) across the genes $j$. For simplicity of notation we shall omit the index $j$ from $\tau_{j}^{-2}$, $\sigma_{j}^{-2}$, $\mathbf{y}_j$ and $\boldsymbol{\beta}_j$ in the remainder of this section. Hence the formulas for $\boldsymbol{\lambda}:=(\boldsymbol{\beta},\tau^{-2},\sigma^{-2})$ below apply to the joint posterior distribution of $(\boldsymbol{\beta}_j, \tau_j^{-2}, \sigma_{j}^{-2})$, for (any) given $j\in\mathcal{J}$.

We shall seek a variational approximation to the posterior distribution of 
$\boldsymbol{\lambda}$ within the class of all distributions with independent marginals
over $\boldsymbol{\beta}$, $\tau^{-2}$ and $\sigma^{-2}$, where we measure the
discrepancy by the Kullback-Leibler (KL) divergence.
Thus letting $p(\boldsymbol{\lambda}|\mathbf{y})$ denote the posterior density
in model \eqref{BSEM}, we seek to find a density $q(\boldsymbol{\lambda})$ of the form
\begin{equation}
q(\boldsymbol{\lambda})
=q_{1}(\boldsymbol{\beta})q_{2}(\tau^{-2})q_{3}(\sigma^{-2}),
\end{equation}
for some marginal densities $q_1, q_2, q_3$, that minimizes the Kullback-Leibler divergence
\begin{equation}
	\label{KL}
	\begin{split}
	\text{KL}(q\vert\vert p) &= \int q(\boldsymbol{\lambda}) \log \frac{q(\boldsymbol{\lambda})}
{p(\boldsymbol{\lambda}|\mathbf{y})}\, d\boldsymbol{\lambda}\\
&= \mathbb{E}_{q} \log q(\boldsymbol{\lambda}) -  \mathbb{E}_{q} \log p(\boldsymbol{\lambda},\mathbf{y}) + \log p(\mathbf{y}), 	
	\end{split}
\end{equation}
over all densities $q$ of product form. Here $p(\mathbf{y})$ denotes the marginal density of
the observation in model \eqref{BSEM}. Because the Kullback-Leibler divergence is nonnegative we have that
\begin{equation}
\label{KL2}
\mathbb{E}_{q} \log p(\boldsymbol{\lambda},\mathbf{y}) - \mathbb{E}_{q} \log q(\boldsymbol{\lambda}) 
\leq \log p(\mathbf{y}).
\end{equation}
Furthermore, minimization of the Kullback-Leibler divergence is equivalent to  maximization of the left side of this
inequality. Thus we may think of the procedure as maximizing a lower bound on the log marginal likelihood.

The solution $q^*$ of this maximization problem, with the marginal densities $q_1, q_2, q_3$ left
completely free, can be seen to be given by densities $q_1^*, q_2^*, q_3^*$ satisying
(see \citep{blei2006,omerod2010})
\begin{equation}
q^\ast_m (\boldsymbol{\lambda}_m) \propto \text{exp}\left\lbrace\displaystyle{\mathbb{E}_{\underset{m^\prime \neq m}{\prod} q_{m^\prime}} \log p(\boldsymbol{\lambda},\mathbf{y})} \right\rbrace, \qquad m=1,2,3.
\end{equation}
In the context of our model this yields 
$q^\ast(\boldsymbol{\lambda}) = q_{1}^\ast(\boldsymbol{\beta})q_{2}^\ast(\tau^{-2})q_{3}^\ast(\sigma^{-2})$,
with the marginal densities (see Section 1 of Supplementary Material (SM)) given by standard distributions,
\begin{equation}
	\label{optVar}
	\begin{aligned}
q_{1}^\ast(\boldsymbol{\beta}) &=^d \mathcal{N}_{p-1}\left(\boldsymbol{\beta}^\ast, \boldsymbol{\Sigma}^\ast \right)\\
q_{2}^\ast(\tau^{-2}) &=^d \mathcal{G}\left(a^\ast, b^\ast \right),\\
q_{3}^\ast(\sigma^{-2}) &=^d \mathcal{G}\left(c^\ast, d^\ast \right),\\
	\end{aligned}
\end{equation}
where the parameters on the right side satisfy 
\begin{equation*}
	\begin{split}
		\boldsymbol{\beta}^\ast& = \displaystyle{ \left( \mathbf{X}^T \mathbf{X} + \mathbb{E}_{q^\ast_2}\left[\tau^{-2}\right] \mathbf{I}_{p-1}\right)^{-1} \mathbf{X}^T \mathbf{y}}\\
		\boldsymbol{\Sigma}^\ast &= \displaystyle{ \left[\mathbb{E}_{q^\ast_3}\left[\sigma^{-2}\right] \left( \mathbf{X}^T \mathbf{X} + \mathbb{E}_{q^\ast_2}\left[\tau^{-2}\right] \mathbf{I}_{p-1}\right) \right]^{-1}}\\
		a^\ast &= \displaystyle{a+\frac{p-1}{2}},\\
		b^\ast &= \displaystyle{b+\frac{1}{2} \mathbb{E}_{q^\ast_3}\left[\sigma^{-2}\right]\mathbb{E}_{q^\ast_1}\left[\boldsymbol{\beta}^T \boldsymbol{\beta}\right]},\\
		c^\ast &= \displaystyle{c+\frac{n+p-1}{2}},\\
		d^\ast &= \displaystyle{d + \frac{1}{2} \mathbb{E}_{q^\ast_1}\left[(\mathbf{y}-\mathbf{X}\boldsymbol{\beta})^T (\mathbf{y}-\mathbf{X}\boldsymbol{\beta})\right] + \frac{1}{2} \mathbb{E}_{q^\ast_2}\left[\tau^{-2}\right] \mathbb{E}_{q^\ast_1}\left[ \boldsymbol{\beta}^T \boldsymbol{\beta}\right]}.
	\end{split}
\end{equation*}
Here $\mathbf{X}$ represents the $n$ by $p-1$ fixed design matrix of \eqref{BSEM}. For the $j^{\text{th}}$ equation in \eqref{BSEM} this is equal to $\mathbf{y}_{-j} = (\mathbf{y}^T_1, \ldots, \mathbf{y}^T_{j-1}, \mathbf{y}^T_{j+1}, \ldots, \mathbf{y}^T_{p})^T$.

Furthermore, the variational lower bound  on the log-marginal likelihood $\log p(\mathbf{y})$ 
(the left side of \eqref{KL2}) evaluated at $q=q^\ast$ simplifies to:
\begin{equation}
	\label{vlb}
	\begin{split}
		\mathcal{L} = &-\frac{n}{2}\log (2\pi) + \frac{1}{2} \log |\boldsymbol{\Sigma}^\ast| + \frac{1}{2} (p-1) + 
		a \log b - \log\Gamma(a) - \\
		& a^\ast \log b^\ast + \log\Gamma(a^\ast) + 
		 c\log d - \log\Gamma(c) - c^\ast\log d^\ast + \\
		 & \log\Gamma(c^\ast) +
		 \frac{1}{2} \mathbb{E}_{q^\ast_3}\left[ \sigma^{-2} \right] \mathbb{E}_{q^\ast_2}\left[\tau^{-2}\right] \mathbb{E}_{q^\ast_1}\left[ \boldsymbol{\beta}^T \boldsymbol{\beta}\right].
	\end{split}
\end{equation}
See SM Section 1 for the details.

The equations \eqref{optVar} express the optimal densities $q_{1}^\ast$, $q_{2}^\ast$ and $q_{3}^\ast$ (or equivalently
the parameters in the right side of \eqref{optVar}) in terms of each other.
This motivates a coordinate ascent algorithm \citep{blei2006, omerod2010} (depicted in Algorithm \ref{AlgVar}), which proceeds by updating 
the parameters in turn, replacing the variational densities on the right hand sides of the equations by their current estimates,
at every iteration. 

Upon convergence the marginal posteriors $p(\boldsymbol{\beta}|\mathbf{y})$, $p(\tau^{-2}|\mathbf{y})$ and $p(\sigma^{-2}|\mathbf{y})$ are approximated by $q_{1}^*(\boldsymbol{\beta})$, $q_{2}^*(\tau^{-2})$ and $q_{3}^\ast(\sigma^{-2})$. Although the algorithm needs to be repeated for each regression equation in \eqref{BSEM}, the overall computational cost of the procedure is low.

In high-dimensional settings with $n<p$ it is preferable to use an SVD decomposition $\mathbf{X}=\mathbf{U}\mathbf{D}\mathbf{V}^T=\mathbf{F}\mathbf{V}^T$ of the regression matrix, and reparameterize the model relative to the $n\times n$ matrix $\mathbf{F}$, i.e.\ in terms of the principal components of $\mathbf{X}^T \mathbf{X}$ \citep{west2003}. Then we compute the variational (Gaussian) approximation $q_1^\ast(\boldsymbol{\theta})$ to the posterior density of the parameter $\boldsymbol{\theta}$ in this reduced model,  and obtain a variational approximation to the marginal posterior density $p(\boldsymbol{\beta}|\mathbf{y})$ of the original parameter by (linearly) transforming back.  For the purpose of edge selection (see Section~\ref{meth:selection}) we are only interested in the posterior expectation $\mathbb{E}_p\left[\boldsymbol{\beta}|\mathbf{y}\right]$ and posterior variance $\mathbb{V}_p\left[\boldsymbol{\beta}|\mathbf{y}\right]$ of $\boldsymbol{\beta}$ (which are approximated by $\mathbb{E}_{q^\ast_1}\left[\boldsymbol{\beta}\right]$ and $\mathbb{V}_{q^\ast_1}\left[\boldsymbol{\beta}\right]$), and hence only the diagonal of the high-dimensional posterior covariance matrix needs to be determined.\\

\begin{algorithm}[H]\footnotesize
\caption{Variational algorithm for local shrinkage}\label{AlgVar}
\begin{algorithmic} [1]
\State \textbf{Initialize:}
\State $b=d=b^{\ast (0)}= d^{\ast (0)} = 0.001$, $\xi=10^{-3}$, $M=1000$ and $t=1$
\While{$|\mathcal{L}^{(t)} - \mathcal{L}^{(t-1)}| \geq \xi$ \textbf{and} $2\leq t \leq M$}
\State update $\displaystyle{\Sigma^{\ast (t)}\gets \left[\mathbb{E}_{q_3^{\ast (t-1)}}(\sigma^{-2}) \left(  \mathbf{X}^T \mathbf{X} + \mathbb{E}_{q_2^{\ast (t-1)}}(\tau^{-2}) \mathbf{I}_{p^\prime} \right)\right]^{-1}}$
\State update $\boldsymbol{\beta}^{\ast (t)} \gets \displaystyle{\mathbb{E}_{q_3^{\ast (t-1)}}(\sigma^{-2}) \Sigma^{\ast (t)} \mathbf{X}^T \mathbf{y}}$
\State update 
\vspace{-15pt}
\begin{equation*}
\begin{split}
d^{\ast (t)} \gets d &+ \displaystyle{\frac{1}{2} \left[(\mathbf{y}-\mathbf{X}\boldsymbol{\beta}^{\ast (t)})^T (\mathbf{y}-\mathbf{X}\boldsymbol{\beta}^{\ast (t)}) + \text{tr}\{ \mathbf{X}^T \mathbf{X} \Sigma^{\ast (t)} \}\right] +} \\
&\quad\displaystyle{\frac{1}{2} \mathbb{E}_{q_2^{\ast (t-1)}}(\tau^{-2}) \left[ {\boldsymbol{\beta}^{\ast (t)}}^T \boldsymbol{\beta}^{\ast (t)} + \text{tr}\{\Sigma^{\ast (t)}\} \right]}
\end{split}
\end{equation*}
\State update $\displaystyle{b^{\ast (t)} \gets b + \frac{1}{2} \mathbb{E}_{q_3^{\ast (t-1)}}(\sigma^{-2}) \left[ {\boldsymbol{\beta}^{\ast (t)}}^T \boldsymbol{\beta}^{\ast (t)} + \text{tr}\{\Sigma^{\ast (t)}\} \right]}$
\State update  $\mathcal{L}^{(t)}$
\State $t \gets t + 1$
\EndWhile
\end{algorithmic}
\end{algorithm}
\null
\vspace{-0.5cm}

%%%%%%%%%%%%%%%%%%%%%%%%%%%%%%%%%%
%%% EMPIRICAL BAYES
%%%%%%%%%%%%%%%%%%%%%%%%%%%%%%%%%%
\subsection{Empirical Bayes and prior calibration}
\label{meth:empBayes}
In the preceding discussion we have treated the vector of hyper-parameters $\boldsymbol{\alpha}=(a,b)$ as fixed.  We now turn to its estimation and present a modified variational algorithm in which $\boldsymbol{\alpha}$ is updated along with the other parameters. The new algorithm is akin to an EM algorithm \citep{braun2010} in which the two steps are, respectively, replaced with a variational E-step, where the lower bound is optimized over the variational parameters via coordinate ascent updates, and a variational M-step, where the lower bound is optimized over $\boldsymbol{\alpha}$ with the variational parameters held fixed.

\begin{figure}[ht]
	\begin{minipage}[c]{0.5\linewidth}
    \centering
  		\subfigure[][Convergence]{
  			\includegraphics[width=1.00\textwidth]{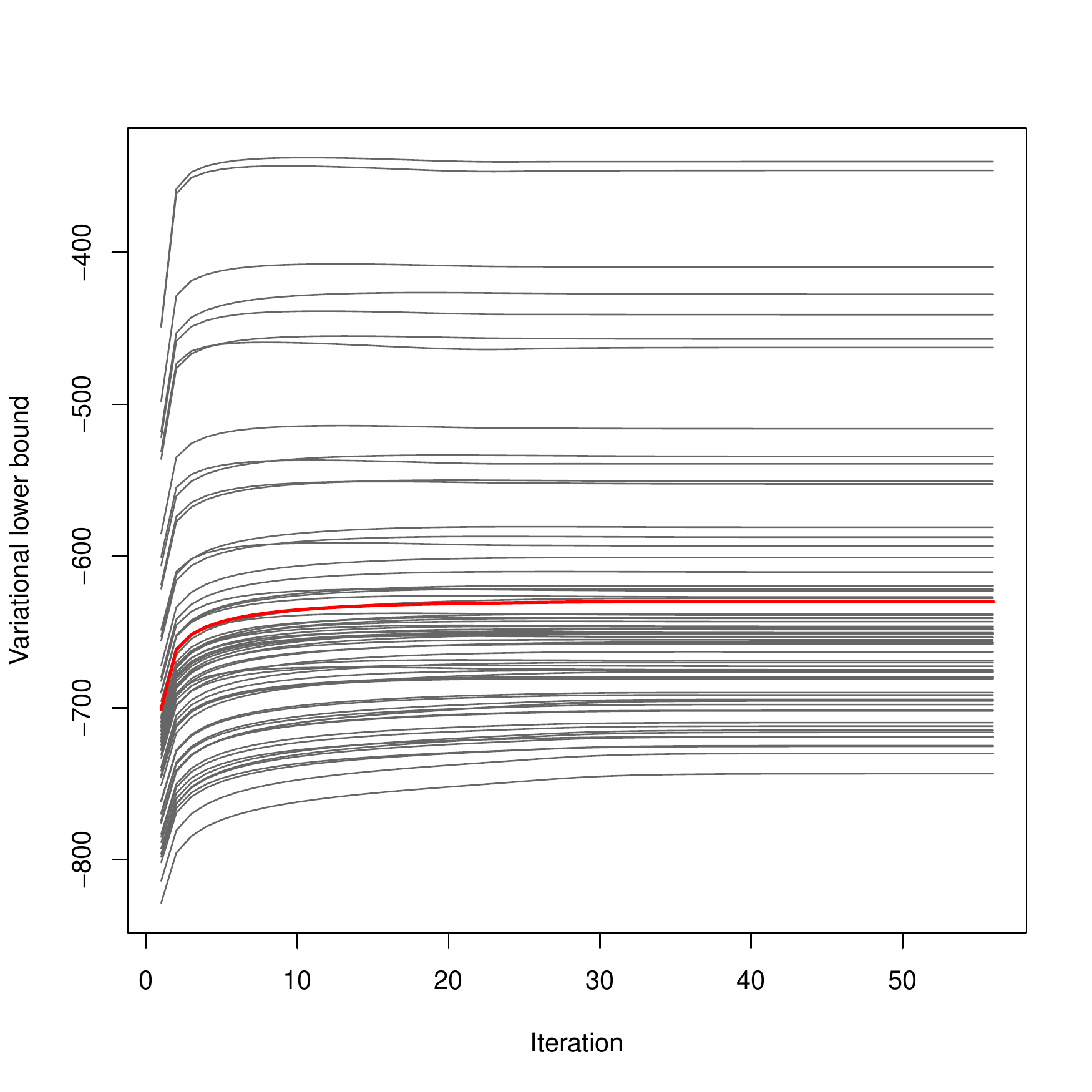}
  			\label{plotVar:sub1}
  		}
  \end{minipage}\hfill
	\begin{minipage}[c]{0.5\linewidth}
    \centering
  		\subfigure[][Global shrinkage prior]{
  			\includegraphics[width=1.00\textwidth]{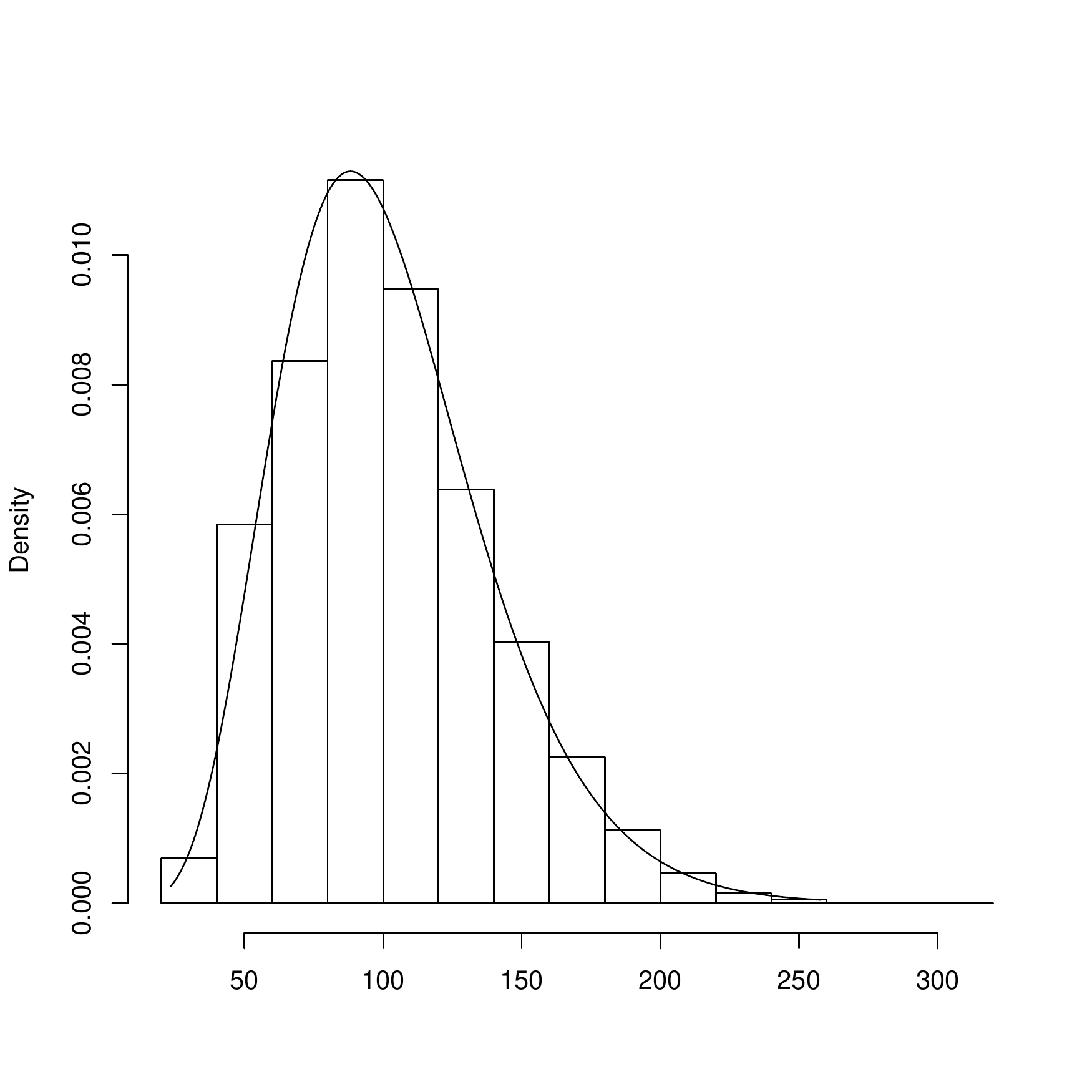}
  			\label{plotVar:sub2}
  		}
  \end{minipage}\hfill
	\caption{Illustration of (a) the convergence of the variational algorithm and (b) the estimated global shrinkage prior on the breast cancer data set (P53 pathway). Figure (a) displays the variational lower bounds $\mathcal{L}_j$ of each regression equation in the SEM as a function of iterations. The red continuous line represents the average lower bound. Figure (b) displays an empirical mixture of marginal posteriors of $\tau_j^{-2}$ obtained by drawing 1000 samples from  $q_2^j(\tau_j^{-2};\mathbf{y}_j)$, $\forall j$. The continuous line represents the density of the estimated global shrinkage prior on $\tau_j^{-2}$, which correspond to $\mathcal{G}(7.404,0.073)$.}
		\label{plotVar}
\end{figure}

We now use the SEM for all genes together, and write the variational
approximation for the posterior density of the parameters for the $j$th gene as $q^{j}$.   (For each $j$ this is given by a triple of three marginal densities.)
The target is to maximize the sum over the genes of the lower bounds 
on the log-marginal likelihood, i.e.\ the sum over $j$ of the left side of \eqref{KL2}, which can be written as
\begin{equation}
\label{LB2}
\sum_{j=1}^{p}{\mathbb{E}_{q^{j }} \log p(\mathbf{y}_j\vert\boldsymbol{\lambda}_j)} 
+ \sum_{j=1}^{p}{\mathbb{E}_{q^{j}} \log \frac{p_{\boldsymbol{\alpha}}(\boldsymbol{\lambda}_j)}{q^{j}(\boldsymbol{\lambda}_j)}} 
\leq \sum_{j=1}^{p}{\log p_{\boldsymbol{\alpha}}(\mathbf{y}_j)}.
\end{equation}
Maximization of the left side with respect to the densities $q^j$ for a fixed hyper-parameter $\boldsymbol{\alpha}$ would lead to the variational estimates $q^{j\ast}$ given by \eqref{optVar}. However, rather than iterating \eqref{optVar} until convergence, we now alternate between ascending in $q$ and in $\boldsymbol{\alpha}$. For the variational estimates $q^{j}$ fixed at their current iterates, optimizing the left-hand side of \eqref{LB2} relative to the parameter $\boldsymbol{\alpha}$ amounts to maximizing, with the current iterate $q^{j\ast}$ replacing $q^j$,
\vspace{-0.5cm}
\begin{equation}
	\begin{split}
		\sum_{j=1}^{p}{\mathbb{E}_{q^{j \ast}} \log p_{\boldsymbol{\alpha}} (\tau_j^{-2})} &= \displaystyle{\sum_{j=1}^{p}{\big(a \log b - \log\Gamma(a)}} \\
		& \qquad + (a-1) \mathbb{E}_{q^{j \ast}}\log\tau_j^{-2} - b \, \mathbb{E}_{q^{j \ast}}\tau_j^{-2}\big).
	\end{split}
\end{equation}
The exact solution to this problem can be found using a fixed-point iteration method, as in \citep{valpola06}. Alternatively, the following approximate 
solution arises by analytical maximization after replacing the digamma function $\psi(x)=\frac{\partial}{\partial x}\log\Gamma(x)$ 
by the approximation $\displaystyle{\log(x)-0.5 x^{-1}}$:
\begin{equation}
\label{sol}
\begin{cases}
	\hat{a}=\displaystyle{\frac{1}{2}\left[ \log\left(\sum_{j=1}^{p}{\mathbb{E}_{q^{j \ast}}\tau_j^{-2}}\right) - p^{-1} \sum_{j=1}^{p}{\mathbb{E}_{q^{j \ast}}\log\tau_j^{-2}}-\log p \right]^{-1}}\\
	\hat{b}=\displaystyle{\hat{a}\cdot p\cdot \left[\sum_{j=1}^{p}{\mathbb{E}_{q^{j \ast}}\tau_j^{-2}}\right]^{-1}}
\end{cases}
\end{equation}
Algorithm~\ref{AlgVar2} outlines how the updates of the hyper-parameters are folded into the variational algorithm. 
At iteration $t$ the hyper-parameters $a^{(t)}$ and $b^{(t)}$ are computed according to \eqref{sol} with the expectations
$\mathbb{E}_{q^{j \ast}}\tau_j^{-2}$ and $\mathbb{E}_{q^{j \ast}}\log\tau_j^{-2}$ computed under the current estimates $q^{j\ast}$. 
Next the variational parameters defining the densities $q^{j\ast}$ are updated according to \eqref{optVar} using the values $a^{(t)}$ and $b^{(t)}$ 
for $a$ and $b$. 
Figure~\ref{plotVar:sub1} illustrates the convergence of the algorithm and shows that the lower bound on the sum of log-marginal likelihoods increases at each step of the algorithm (red line). Although this is not true for the lower bounds of each regression equation in the SEM, this does demonstrate that the estimation procedure yields a well-informed prior that is beneficial overall.

The second summand on the left-hand side of \eqref{LB2} is equal to minus
$\sum_{j=1}^{p} \text{KL}(q^{j \ast} \vert\vert p_{\boldsymbol{\alpha}})$. This suggests that the procedure will seek to set the hyper parameters $\boldsymbol{\alpha}$ so that the prior density
$p_{\boldsymbol{\alpha}}$ of the $\boldsymbol{\lambda}_j$ on the average most resembles their (approximate) posteriors $q^{j\ast}$, based on the different genes. This connects to the recent work of \citet{vanDeWiel2013} on shrinkage priors for differential gene expression analysis, whose empirical Bayes procedure consists in finding $\boldsymbol{\alpha}$ such that $p_{\boldsymbol{\alpha}}(\tau_j^{-2})\approx n^{-1} \sum_j{p_{\boldsymbol{\alpha}}(\tau_j^{-2}\vert
  \mathbf{y}_j)}$.
Figure \ref{plotVar:sub2} shows that our approach fulfills the same objective. It is natural for the empirical Bayes procedure to have this ``averaging of marginal posteriors'' property, as it attempts to calibrate the prior according to the data. The role of the global shrinkage prior $\mathcal{G}(a,b)$ is to encourage the posterior distributions of the $\tau_j^{-2}$, for $j\in\mathcal{J}$, to shrink to a common distribution, centered around the (prior) mean $a/b$.\\

\begin{algorithm}[h]\footnotesize
\caption{ Variational EM algorithm with global-local shrinkage priors}\label{AlgVar2}
\begin{algorithmic} [1]
\State \textbf{Initialize:}
\State $a^{(0)}= b^{(0)}=a^{\ast (0)}= 0.001, \forall j\in\mathcal{J},\ b_j^{\ast (0)}= d_j^{\ast (0)} = 0.001$, $\xi=10^{-3}$, $M=1000$ and $t=1$
\While{$\text{max}|\mathcal{L}_j^{(t)} - \mathcal{L}_j^{(t-1)}| \geq \xi$ \textbf{and} $2\leq t \leq M$}
\Statex \hspace{\algorithmicindent}\texttt{E-step: Update variational parameters:}
\For{$j=1$ to $p$}
\State update $\Sigma_j^{\ast (t)}$, $\boldsymbol{\beta}_j^{\ast (t)}$, $d_j^{\ast (t)}$, $b_j^{\ast (t)}$ and $\mathcal{L}_j^{(t)}$ in that order (as in \textbf{Algorithm 1})
\State update $a^{\ast (t)} \gets a^{(t-1)} + \frac{p-1}{2}$
\EndFor
%\null\\
\Statex \hspace{\algorithmicindent}\texttt{M-step: Update hyper-parameters:}
\State $\displaystyle{a^{(t)} \gets 0.5 \left( p^{-1}\sum_{j=1}^{p}{\left(\log(b_j^{\ast (t)})-\psi(a^{\ast (t)})\right)} - \log p + \log \sum_{j=1}^{p}{\frac{a^{\ast (t)}}{b_j^{\ast (t)}}} \right)^{-1}}$
\State $\displaystyle{b^{(t)} \gets a^{(t)}\cdot p \left(\sum_{j=1}^{p}{\frac{a^{\ast (t)}}{b_j^{\ast (t)}}}\right)^{-1}}$
\State $t \gets t + 1$
\EndWhile
\end{algorithmic}
\end{algorithm}
\null\vspace{-0.5cm}

%%%%%%%%%%%%%%%%%%%%%%%%%%%%%%%%%%
%%% POSTERIOR THRESHOLDING
%%%%%%%%%%%%%%%%%%%%%%%%%%%%%%%%%%
\subsection{Edge selection}
\label{meth:selection}

In this section we describe a separate procedure for edge selection. 
This consists of first ranking the edges based on summary statistics from the (marginal) posterior distributions under the model \eqref{BSEM} obtained in the preceding sections, and next performing forward selection along this ordering. For the latter we use Bayes factors and their relation to a Bayesian version of the local false discovery rate
\citep[lfdr]{efron2010}.

\subsubsection{Edge ordering}
\label{meth:selection:order}
Denote the approximate posterior expectation and variance of $\beta_{j,k}$ obtained in Sections~\ref{meth:VB} and~\ref{meth:empBayes} for SEM \eqref{BSEM} by $\mathbb{E}_{q^{j \ast}}\left[\beta_{j,k}|\mathbf{y}_j\right]$ and $\mathbb{V}_{q^{j \ast}}\left[\beta_{j,k}|\mathbf{y}_j\right]$, and define
\begin{equation}
	\label{kappa}
	\kappa_{j,k}=\frac{\left\lvert \mathbb{E}_{q^{j \ast}}\left[\beta_{j,k}|\mathbf{y}_j\right]\right\rvert }{ \sqrt{\mathbb{V}_{q^{j \ast}}\left[\beta_{j,k}|\mathbf{y}_j\right]}},\qquad  j,k\in\mathcal{J}\ \text{with}\ j\neq k.
\end{equation}
Next for a given edge $(j,k)$ (between genes $j$ and $k$) define the quantity $\bar{\kappa}_{j,k} = (\kappa_{j,k}+\kappa_{k,j})/2$, and order the set of $P=p(p-1)/2$ edges according to their associated values $\bar{\kappa}_{j,k}$, from large to small.  Let $(j(r),k(r))$ denote the $r$th edge in this ordering, and abbreviate its associated value to $\bar{\kappa}_{r}= \bar{\kappa}_{j(r),k(r)}$. This ordering is
retained in all of the following. However, we do not necessarily select all edges below a certain threshold, but proceed by forward selection, for $r=1,\ldots,P$.

\subsubsection{Bayes factors}
\label{meth:selection:BF}
Selection at stage $r$ (see Section \ref{meth:selection:sel}) will be based on Bayes factors $\text{BF}(j(r),k(r))$ and $\text{BF}(k(r),j(r))$ for the two regression parameters $\beta_{j(r),k(r)}$ and $\beta_{k(r),j(r)}$ associated with the $r$th edge.

Denote by $m_{j(r),k(r),1}$ the model in SEM \eqref{BSEM} for the response variable $\mathbf{y}_{j(r)}$, with the covariates (or nonzero $\beta_{j(r),k}$) restricted to the edge $(j(r),k(r))$ and any \emph{previously  selected} edge (involving node $j(r)$) with rank lower or equal to $r-1$. Likewise, define $m_{j(r), k(r),0}$, but with the restriction $\beta_{j(r),k(r)}=0$, which is equivalent to the exclusion of edge $(j(r),k(r))$. The Bayes factor associated with this model is
\begin{equation}
	\label{BF}
	\text{BF}(j(r),k(r)) = \frac{p(\mathbf{y}_{j(r)}|m_{j(r),k(r),1})}{p(\mathbf{y}_{j(r)}|m_{j(r),k(r),0})},\qquad r=1,\ldots,P.
	% \text{ML}_{[r],1}/\text{ML}_{[r],0}.
\end{equation}
The Bayes factor $\text{BF}(k(r),j(r))$ is defined analogously from the regression models $m_{k(r),j(r),1}$ and $m_{k(r),j(r),0}$ for response
variable $\mathbf{y}_{j(k)}$.

\subsubsection{Prior for Bayesian variable selection}
\label{meth:selection:prior}
The global shrinkage prior for the precision parameters $\tau^{-2}_j$ estimated from the data in Section~\ref{meth:empBayes} is not appropriate for computing the Bayes factors \eqref{BF}. Because it has been calibrated (by the variational Bayes method outlined in Algorithm~\ref{AlgVar2}) for the network comprised of all edges, it is likely to be located away from zero, which will induce strong regularization on the regression parameters, making it difficult for the Bayes factors to discriminate between the subsequent models (in particular when $n$ is small). A non-informative prior runs into the same problem (perhaps even in a more sever manner).

Motivated by the Zellner-Siow prior \citep{liang2013,zellner1980} we propose to employ instead the ``default prior'' $\tau^{-2}_j\sim\mathcal{G}(1/2,n/2)$. This concentrates near its prior expectation $n^{-1}$  (i.e. the fixed unit information prior of \citet{kass1995}), and hence is concentrated near $0$ for moderate and large values of $n$, while less stringent for small $n$ (see illustration in SM Section 4).

\subsubsection{Bayesian analogue of lfdr}
\label{meth:selection:blfdr}
Since both Bayes factors $\text{BF}(j(r),k(r))$ and $\text{BF}(k(r),j(r))$ are informative for the
relevance of edge $(j(r),k(r))$, we need to combine these and find a suitable threshold. For that
purpose, we link the Bayes factors to the posterior null-probability
$\text{P}_0(\bar{\kappa}_{r}) = P(\beta_{j(r),k(r)}=0,\beta_{k(r),j(r)}=0 | \mathbf{y})$, where
$\mathbf{y} = (\mathbf{y}^T_1, \ldots, \mathbf{y}^T_{p})^T$. The absence of edge $(j(r),k(r))$ is
reflected by $\beta_{j(r),k(r)}=\beta_{k(r),j(r)}=0$, which, in the spirit of forward
selection, implies the null models $m_{j(r), k(r),0}$ and $m_{k(r),j(r),0}$. The posterior
null-probability is linked to the local false discovery rate \citep[lfdr]{efron2010}. However, as in
\citet{vanDeWiel2013}, we condition on the data $\mathbf{y}$ rather than on a test statistic. Then,
we have
\begin{equation}
\begin{split}
	\label{lfdr}
	\text{P}_0(\bar{\kappa}_{r}) &= P(\beta_{j(r),k(r)}=0,\beta_{k(r),j(r)}=0 | \mathbf{y})\\ &\leq \min \{P(\beta_{j(r),k(r)}=0 | \mathbf{y}), P(\beta_{k(r),j(r)}=0 | \mathbf{y})\}.
\end{split}\end{equation}
Here, the bound is used because the SEM may not provide accurate joint probabilities on regression coefficients from different regression models.
Now, assume the prior null probability $P(\beta_{j,k}=0 | \mathbf{y}_{-j})=p_0$, $\forall j\in\mathcal{J}$ , where $\mathbf{y}_{-j} = (\mathbf{y}^T_1, \ldots, \mathbf{y}^T_{j-1}, \mathbf{y}^T_{j+1}, \ldots, \mathbf{y}^T_{p})^T$. Note that a constant value of $p_0$ is reasonable, because it simply reflects the prior probability that response $\mathbf{y}_{j}$ does not respond to covariate $\mathbf{y}_{k}$ (which is a member of $\mathbf{y}_{-j}$). Then,
\begin{equation}
\begin{split}
	\label{lfdr2}
	P(\beta_{j,k}=0 | \mathbf{y}) &= P(\beta_{j,k}=0 | \mathbf{y}_j, \mathbf{y}_{-j}) \\
	&=  \frac{P(\mathbf{y}_j | \beta_{j,k}=0, \mathbf{y}_{-j})P(\beta_{j,k}=0 | \mathbf{y}_{-j})}{P(\mathbf{y}_j | \mathbf{y}_{-j})}\\
	&= \frac{P(\mathbf{y}_j | m_{j,k,0})p_0}{P(\mathbf{y}_j | m_{j,k,0})p_0 + (1-p_0)P(\mathbf{y}_j | m_{j,k,1})} \\
	&= \frac{p_0}{p_0 + (1-p_0)\text{BF}(j,k)}.
\end{split}
\end{equation}
Define the max Bayes factor: $\text{BF}(\bar{\kappa}_{r}) = \max \{\text{BF}(j(r),k(r)),\text{BF}(k(r),j(r)\}$. Then, after substituting (\ref{lfdr2}) into (\ref{lfdr}) we have, for threshold $\gamma = (1-\alpha)p_0/(\alpha(1-p_0))$,
\begin{equation}
	\label{rule}
	\text{BF}(\bar{\kappa}_{r}) \geq \gamma \Longleftrightarrow \text{P}_0(\bar{\kappa}_{r}) \leq \alpha.
\end{equation}
Equation \eqref{rule} suggests that edges in the graph can be selected using a thresholding rule on the Bayes factors that controls the posterior null-probability. For example, when we have $p_0=0.9$, then $\text{BF}(\bar{\kappa}_{r})>81$  implies $\text{P}_0(\bar{\kappa}_{r})<0.1$. However, to use this approach an estimate of $p_0$ is required. We simply propose
\begin{equation}
	\label{p0}
	\hat{p}_0 = \frac1{2P}\biggl(\sum_{r=1}^{P} (I_{\{\text{BF}'(j(r),k(r)) \leq 1\}} + I_{\{\text{BF}'(k(r),j(r)) \leq 1\}})\biggr).
\end{equation}
where $\text{BF}'(j(r),k(r))$ is defined analogously to $\text{BF}(j(r),k(r))$, but \emph{without} forward selection (so all covariates corresponding to edge ranks $\leq r$ are included), because the forward selection procedure requires knowing $\hat{p}_0$.

\subsubsection{Forward selection procedure}
\label{meth:selection:sel}
We introduce the following sequential procedure to update the set $\mathsf{E}$ of selected edges 
and the models $m_{j(r), k(r),0}$, $m_{j(r), k(r),1}$, $m_{k(r),j(r),0}$, $m_{k(r),j(r),1}$ when increasing $r$:
\begin{enumerate}
	\item Initiate $\alpha$, $r=1$, $\ell=0$ and $\mathsf{E}^{0}=\emptyset$. Compute $\gamma$ from $\alpha$ and $\hat{p}_0$.
    \item Determine the models $m_{j(r), k(r),0}$ and $ m_{k(r),j(r),0}$ which are the current models for $\mathbf{y}_{j(r)}$ and $\mathbf{y}_{k(r)}$ that correspond to
    $\mathsf{E}^{r-1}$. Augment
    those models with covariates $\mathbf{y}_{k(r)}$ and $\mathbf{y}_{j(r)}$, respectively, and fit these models to obtain $m_{j(r), k(r),1}$ and $ m_{k(r),j(r),1}$.
	\item Calculate the max Bayes factor $\text{BF}(\bar{\kappa}_{r})$
	\item Only if $\text{BF}(\bar{\kappa}_{r})>\gamma$ update $\mathsf{E}^{r}=\mathsf{E}^{r-1}\cup\{(j(r),k(r))\}$
	\item Update $r=r+1$ and go back to step 2
\end{enumerate}
For the purpose of variable selection we include intercepts in the SEM. Finally, we estimate $\mathcal{E}$ by the last update of $\mathsf{E}$.

The selection procedure respects the initial ranking of the edges in terms of the order in which they are considered for inclusion in the forward selection. However, the procedure is set up to proceed when a given edge is not selected, because in the light of the current model subsequent edges may (slightly) increase the marginal likelihood. As in practice we observed that the Bayes factor decreases with $r$ (see Supplementary Figure 2), a stopping criterion
may be practical if  $P$ is large; e.g.\ stop if $r$ reaches $r_{\text{max}} = (1-\hat{p}_0) P$, or if $\text{BF}(\bar{\kappa}_{r})$ has not exceeded $\gamma$ for, say, 100 consecutive values of $r$.

%%%%%%%%%%%%%%%%%%%%%%%%%%%%%%%%%%%%%%%%%%%%%%%%%%%%%%
%%%%%%%%%%%%%%%%%%%%%%%%%%%%%%%%%%%%%%%%%%%%%%%%%%%%%%
%%%	section: MODEL-BASED SIMULATION
%%%%%%%%%%%%%%%%%%%%%%%%%%%%%%%%%%%%%%%%%%%%%%%%%%%%%%
%%%%%%%%%%%%%%%%%%%%%%%%%%%%%%%%%%%%%%%%%%%%%%%%%%%%%%

\section{Model-based simulation}
\label{modelSim}
\begin{figure}[h]
	\begin{minipage}[c]{0.25\linewidth}
    \centering
  		\subfigure[][Band]{
  			\includegraphics[width=1.00\textwidth]{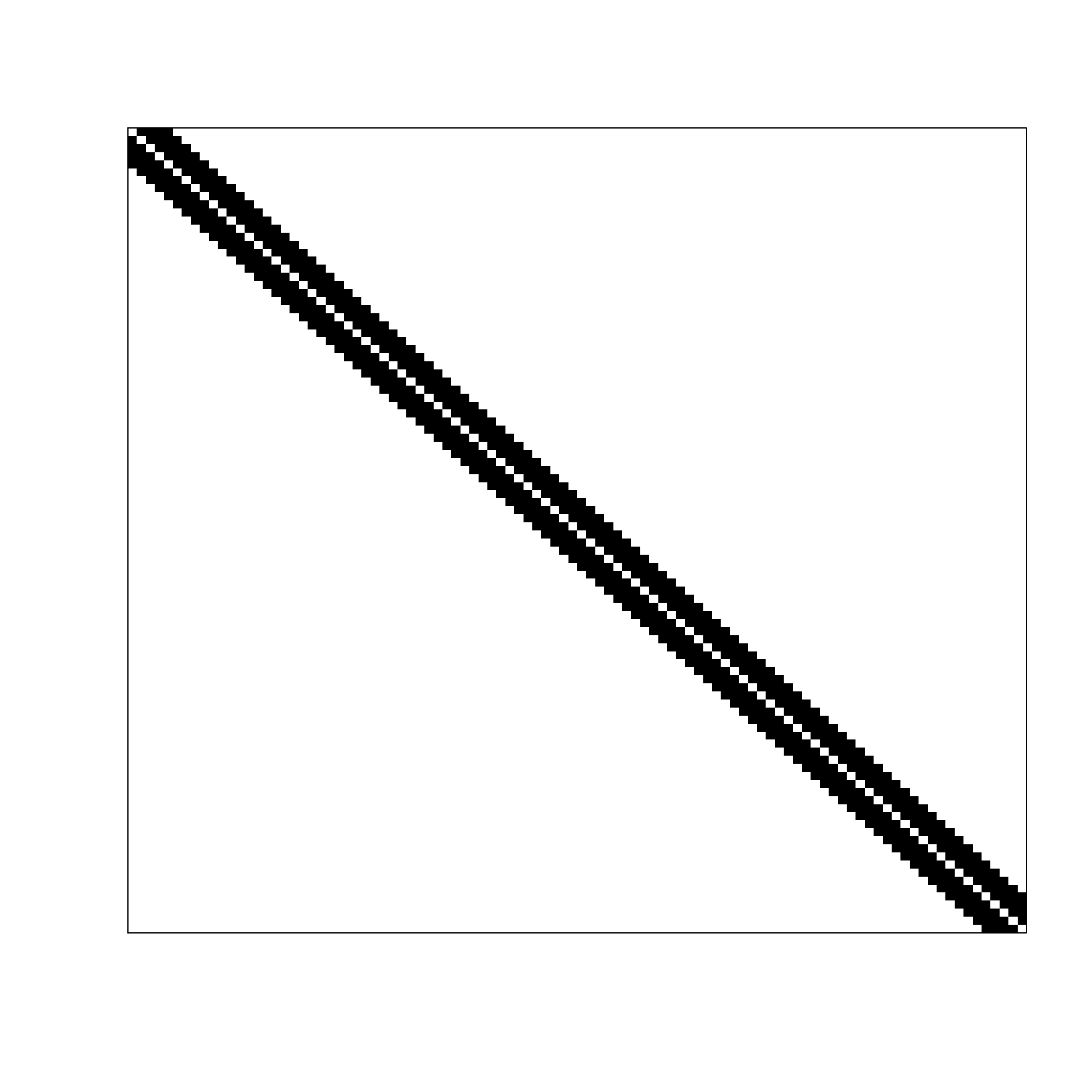}
  			\label{structure:sub1}
  		}
  \end{minipage}\hfill
	\begin{minipage}[c]{0.25\linewidth}
    \centering
  		\subfigure[][Cluster]{
  			\includegraphics[width=1.00\textwidth]{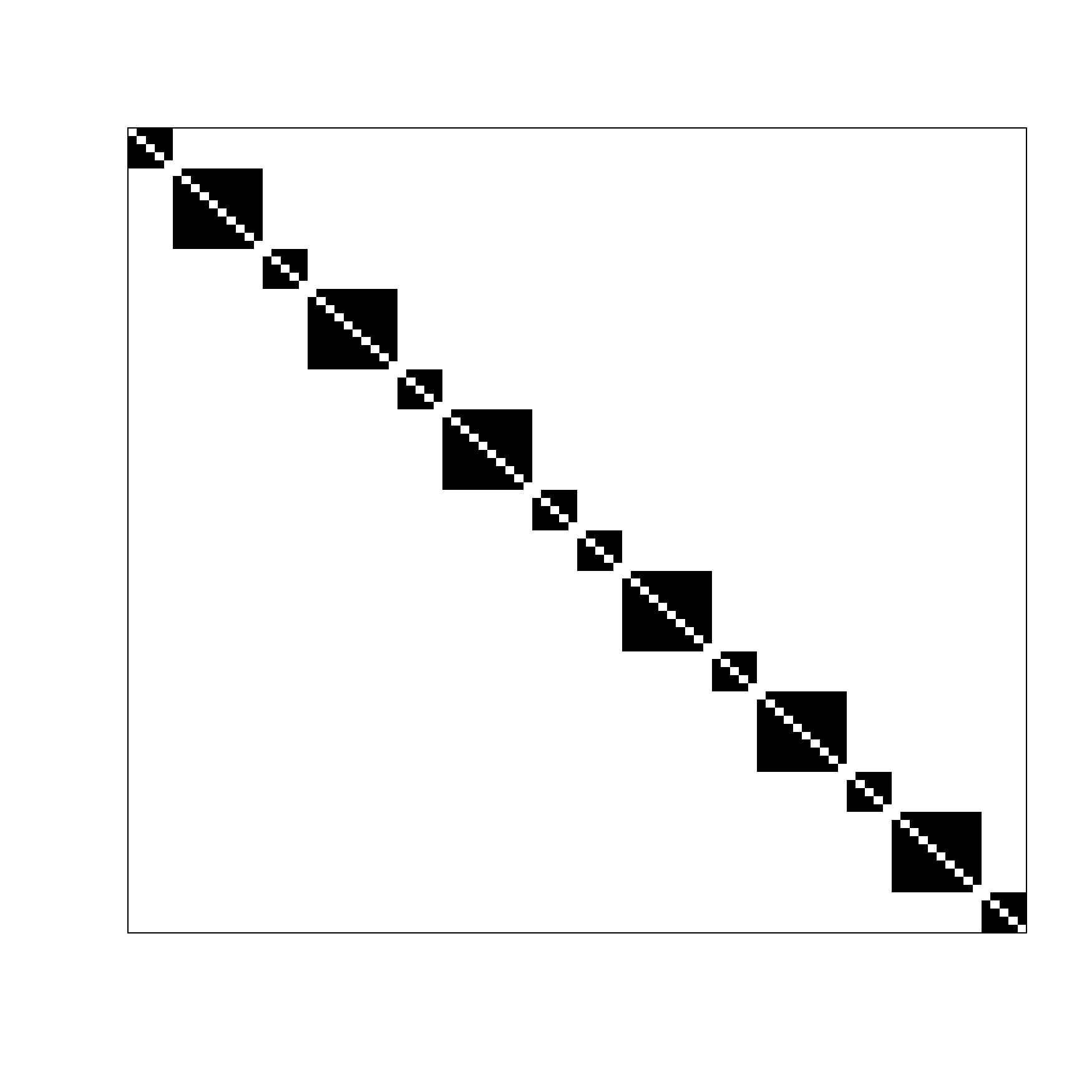}
  			\label{structure:sub2}
  		}
  \end{minipage}\hfill
  \begin{minipage}[c]{0.25\linewidth}
    \centering
  		\subfigure[][Hub]{
  			\includegraphics[width=1.00\textwidth]{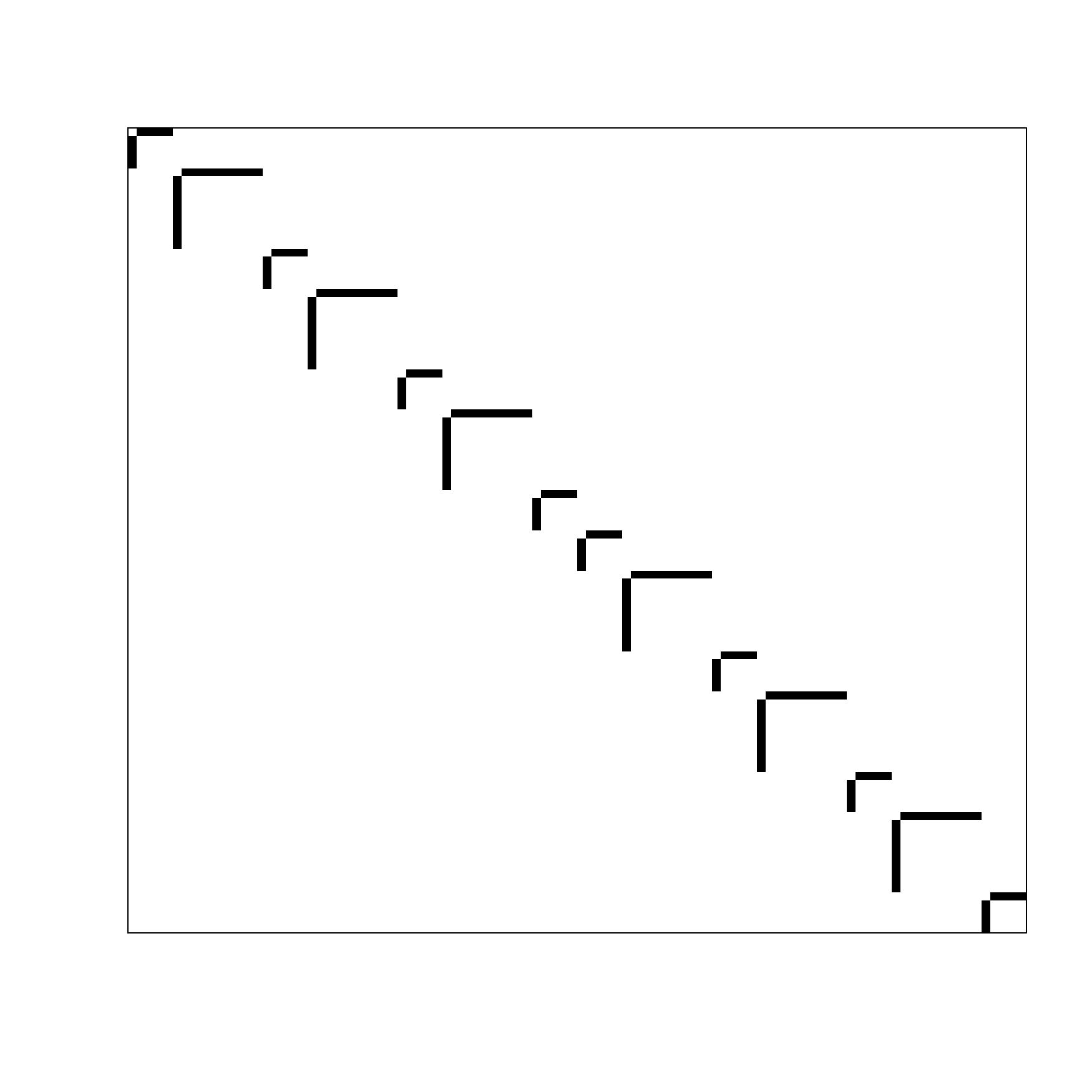}
  			\label{structure:sub3}
  		}
  \end{minipage}\hfill
  \begin{minipage}[c]{0.25\linewidth}
    \centering
  		\subfigure[][Random]{
  			\includegraphics[width=1.00\textwidth]{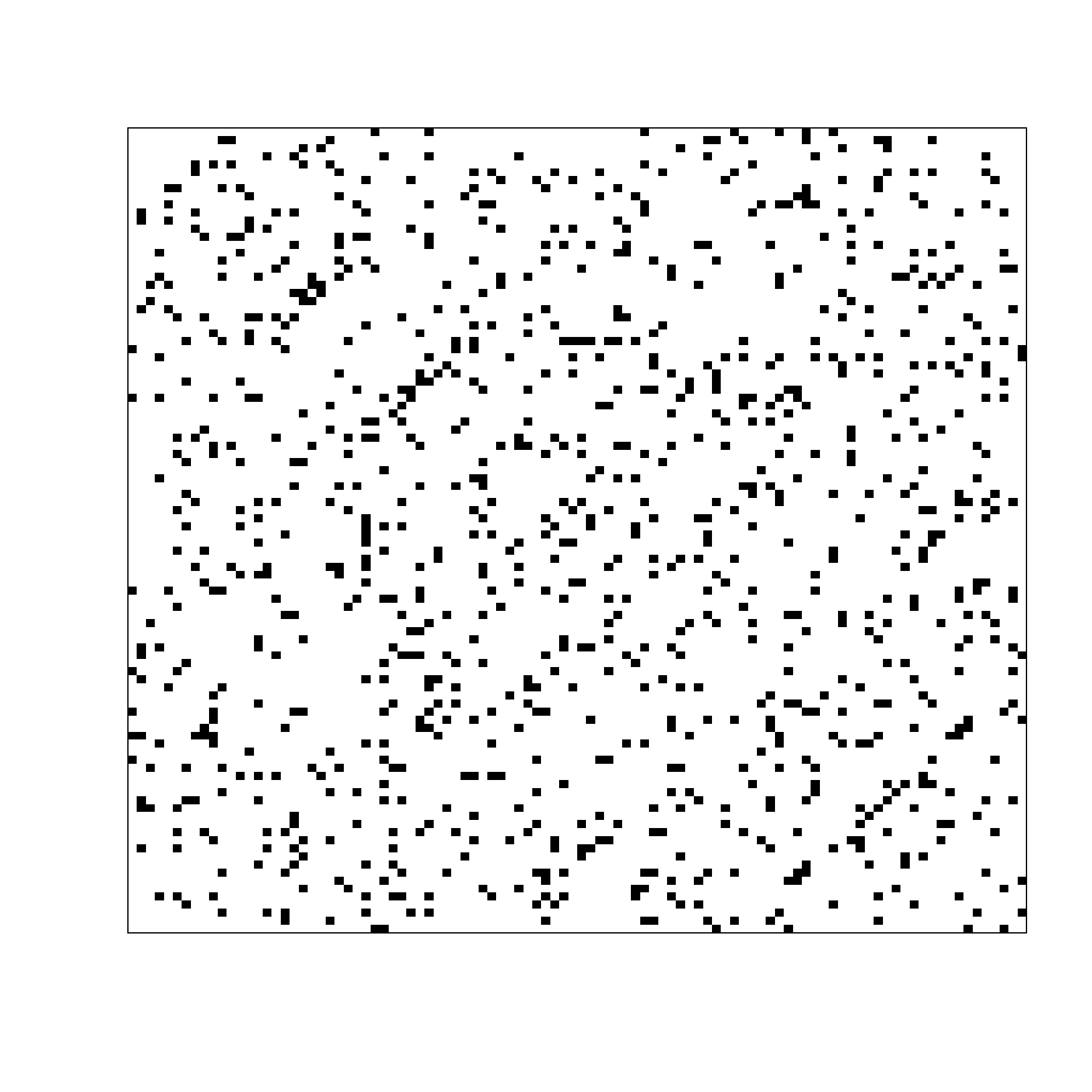}
  			\label{structure:sub4}
  		}
  \end{minipage}\hfill
		\caption{Graph structures considered for the precision matrix $\boldsymbol{\Omega}$ in our simulation. Black and white dots represent non-zero and zero entries, respectively. Only off-diagonal elements are displayed. For precision matrices with block-diagonal structures (clusters and hubs), block sizes were set to 5 and 10. In (a) the bandwidth is equal to four. The graph density $\delta$ is (a) $\delta=0.079$, (b) $\delta=0.071$, (c) $\delta=0.017$ and (d) $\delta=0.096$.}
		\label{structure}
\end{figure}

In this section we investigate the performance of our approach, termed ShrinkNet, in recovering the structure of an undirected network and compare it to popular approaches. We generate $n\in \{25,50,100\}$ samples from a multivariate normal distribution with mean vector $\mathbf{0}$ and $100\times 100$ precision matrix $\boldsymbol{\Omega}$, corresponding to four different graph structures: \textit{band}, \textit{cluster}, \textit{hub} and \textit{random} \citep{zhao2012} (see Figure~\ref{structure} for illustration), every of them sparse, with graph density ranging from $0.017$ to $0.096$. We generated the inverse covariance matrix $\boldsymbol{\Omega}$ corresponding to each graph structure from a G-Wishart distribution \citep{mohammadi2015} with scale matrix equal to the identity and $b=4$ degrees of freedom. In SM Section~2 we provide statistical summaries on the magnitude of the generated partial correlations.

We compared our approach ShrinkNet to the popular frequentist SEM with the Lasso penalty (SEM$_\text{L}$) \citep{meinshausen2006}, the Graphical Lasso (GL$_{\lambda}$) \citep{friedman2008}, and GeneNet \citep{schafer2006}. The latter combines a non-sparse linear shrinkage estimator with an \textit{a posteriori} edge selection procedure. For the purpose of comparison with ShrinkNet, we also consider the Bayesian SEM \eqref{BSEM} with the non-informative global shrinkage prior
$\mathcal{G}(0.001,0.001)$, which we subsequently refer to as `NoShrink'.

Briefly, graph selection is as follows. For SEM$_\text{L}$ and GL$_{\lambda}$ we use the EBIC criterion \citep{foygel2010,chen2008} for selecting the optimal regularization parameter(s), whereas for GeneNet and ShrinkNet a threshold of $0.1$ on the local false discovery rate and the posterior null probability $\text{P}_0$ is employed. In SM Section 3 we provide more details as to how an edge ranking is obtained for each method.

To evaluate the performance of the methods in recovering the graph structures we report partial ROC curves (SM Section 5), which depict the true positive rate (TPR) as a function of the false positive rate (FPR) for FPR$<0.2$), and various performance measures on selected graphs. Table \ref{perfMeasure} below provides a numerical summary: it displays averages and standard deviations of TPRs, FPRs, F-scores and partial AUCs (pAUC) \citep{dodd2003} as a function of the method, $n$ and the true graph structure. The F-score=$2\times (\text{precision}\times \text{TPR})/(\text{precision} + \text{TPR})$ is a popular performance measure, defined as the harmonic mean between the TPR=TP/(TP+FN) (also called \textit{recall}) and the precision=TP/(TP+FP), where TP, FP, and FN are the number of true positives, false positives, and false negatives, respectively.

Table~\ref{perfMeasure} shows that ShrinkNet achieves the highest average partial AUCs in all but two situations in which it is very close to the maximum. The results also indicate that NoShrink is often outperformed by GeneNet, and comparable to GL$_{\lambda}$, which suggests that the global shrinkage carried out by ShrinkNet considerably improves edge ranking. SEM$_\text{L}$ has the lowest average pAUC in all situations.

The performance of each method in recovering the true graph structure can also be evaluated by the average F-score. According to this metric the best performance is achieved by NoShrink and ShrinkNet in all but two cases. In moderate- ($n=50$) and high-dimensional cases ($n=25$), NoShrink and ShrinkNet show a much larger F-score than others. This is particularly pronounced when $n=25$, in which case GL$_{\lambda}$ and GeneNet have an F-score (and TPR) very close to zero. In this context SEM$_\text{L}$ is performing better than GL$_{\lambda}$ and GeneNet, but worse than NoShrink and ShrinkNet.

%%%% BAND
\begin{table}\fontsize{7.50pt}{7.5pt}\selectfont%\tiny
\label{perfMeasure}
\centering
\begin{tabular}{r|l|l|ccc|c}
  \hline\hline
& n & Method & TPR & FPR & F-score & pAUC \\ 
  \hline
%%%%%%%%%%%%%%%%
%%%% BAND
%%%%%%%%%%%%%%%%
\multirow{15}{*}{\textbf{\rotatebox{90}{BAND}}} & \multirow{5}{*}{$100$} & NoShrink & 0.255 (0.014) & 0.001 (0.000) & \textbf{0.404 (0.017)} & 0.342 (0.007) \\ 
& &  ShrinkNet & 0.246 (0.015) & 0.001 (0.000) & 0.392 (0.020) & \textbf{0.354 (0.007)} \\ 
 & & SEM$_\text{L}$ & 0.143 (0.013) & 0.000 (0.000) & 0.249 (0.020) & 0.331 (0.006) \\ 
 & & GL$_{\lambda}$ & 0.144 (0.065) & 0.001 (0.001) & 0.242 (0.094) & 0.334 (0.007) \\ 
 & & GeneNet & 0.226 (0.025) & 0.001 (0.000) & 0.366 (0.032) & 0.351 (0.007) \\ \cline{2-7} 
 & \multirow{5}{*}{$50$} & NoShrink & 0.150 (0.015) & 0.001 (0.001) & \textbf{0.257 (0.022)} & 0.292 (0.009) \\ 
 & & ShrinkNet & 0.142 (0.013) & 0.001 (0.001) & 0.245 (0.020) & \textbf{0.312 (0.010)} \\ 
 & & SEM$_\text{L}$ & 0.059 (0.011) & 0.000 (0.000) & 0.110 (0.019) & 0.273 (0.010) \\ 
  & & GL$_{\lambda}$ & 0.013 (0.010) & 0.000 (0.000) & 0.028 (0.018) & 0.294 (0.009) \\ 
  & & GeneNet & 0.084 (0.020) & 0.000 (0.000) & 0.154 (0.034) & 0.309 (0.009) \\  \cline{2-7} 
 & \multirow{5}{*}{$25$} & NoShrink & 0.065 (0.011) & 0.002 (0.001) & 0.119 (0.019) & 0.244 (0.010) \\ 
 & & ShrinkNet & 0.066 (0.011) & 0.002 (0.001) & \textbf{0.122 (0.020)} & 0.259 (0.010) \\ 
 & & SEM$_\text{L}$ & 0.017 (0.007) & 0.000 (0.000) & 0.033 (0.014) & 0.220 (0.008) \\ 
 & & GL$_{\lambda}$ & 0.004 (0.004) & 0.000 (0.000) & 0.011 (0.006) & 0.247 (0.009) \\ 
  & & GeneNet & 0.000 (0.002) & 0.000 (0.000) & 0.011 (0.008) & \textbf{0.260 (0.010)} \\   
%%%%%%%%%%%%%%%%
%%%% CLUSTER
%%%%%%%%%%%%%%%%
  \hline\hline
\multirow{15}{*}{\textbf{\rotatebox{90}{CLUSTER}}}& \multirow{5}{*}{$100$} & NoShrink & 0.290 (0.015) & 0.000 (0.000) & 0.449 (0.018) & 0.375 (0.008) \\ 
   & & ShrinkNet & 0.276 (0.015) & 0.000 (0.000) & 0.432 (0.018) & \textbf{0.389 (0.007)} \\ 
   & & SEM$_\text{L}$ & 0.175 (0.014) & 0.000 (0.000) & 0.297 (0.020) & 0.336 (0.007) \\ 
   & & GL$_{\lambda}$ & 0.402 (0.038) & 0.001 (0.001) & 0.566 (0.037) & 0.357 (0.007) \\ 
   & & GeneNet & 0.299 (0.027) & 0.000 (0.000) & \textbf{0.458 (0.031)} & 0.383 (0.008) \\  \cline{2-7} 
   & \multirow{5}{*}{$50$} & NoShrink & 0.174 (0.016) & 0.000 (0.000) & \textbf{0.295 (0.022)} & 0.326 (0.010) \\ 
   & & ShrinkNet & 0.163 (0.015) & 0.000 (0.000) & 0.279 (0.022) & \textbf{0.350 (0.010)} \\ 
   & & SEM$_\text{L}$ & 0.078 (0.010) & 0.000 (0.000) & 0.144 (0.018) & 0.280 (0.008) \\ 
   & & GL$_{\lambda}$ & 0.022 (0.013) & 0.000 (0.000) & 0.043 (0.025) & 0.317 (0.008) \\ 
   & & GeneNet & 0.120 (0.026) & 0.000 (0.000) & 0.213 (0.040) & 0.344 (0.009) \\  \cline{2-7} 
  &\multirow{5}{*}{$25$} & NoShrink & 0.084 (0.013) & 0.001 (0.001) & 0.153 (0.021) & 0.275 (0.010) \\ 
  &  & ShrinkNet & 0.086 (0.014) & 0.001 (0.001) & \textbf{0.156 (0.023)} & \textbf{0.295 (0.011)} \\ 
   & & SEM$_\text{L}$ & 0.031 (0.008) & 0.000 (0.000) & 0.059 (0.015) & 0.228 (0.009) \\ 
   & & GL$_{\lambda}$ & 0.004 (0.005) & 0.000 (0.000) & 0.013 (0.009) & 0.270 (0.008) \\ 
   & & GeneNet & 0.004 (0.006) & 0.000 (0.000) & 0.017 (0.013) & 0.293 (0.010) \\    
%%%%%%%%%%%%%%%%
%%%% Hub
%%%%%%%%%%%%%%%%
  \hline\hline
\multirow{15}{*}{\textbf{\rotatebox{90}{HUB}}}& \multirow{5}{*}{$100$} & NoShrink & 0.357 (0.031) & 0.000 (0.000) & 0.520 (0.035) & 0.385 (0.014) \\ 
  &  & ShrinkNet & 0.354 (0.030) & 0.000 (0.000) & 0.519 (0.034) & \textbf{0.388 (0.013)} \\ 
  & & SEM$_\text{L}$ & 0.277 (0.027) & 0.000 (0.000) & 0.431 (0.033) & 0.355 (0.013) \\ 
  & & GL$_{\lambda}$ & 0.459 (0.055) & 0.004 (0.001) & \textbf{0.540 (0.035)} & 0.384 (0.014) \\ 
  & & GeneNet & 0.302 (0.043) & 0.001 (0.001) & 0.446 (0.047) & 0.381 (0.013) \\ \cline{2-7} 
  &  \multirow{5}{*}{$50$} & NoShrink & 0.241 (0.030) & 0.001 (0.000) & 0.378 (0.039) & 0.346 (0.016) \\ 
  &  & ShrinkNet & 0.242 (0.029) & 0.000 (0.000) & \textbf{0.383 (0.038)} & \textbf{0.353 (0.015)} \\ 
  & & SEM$_\text{L}$ & 0.149 (0.026) & 0.000 (0.000) & 0.258 (0.038) & 0.310 (0.017) \\ 
  & & GL$_{\lambda}$ & 0.183 (0.052) & 0.001 (0.001) & 0.295 (0.067) & 0.345 (0.014) \\ 
  & & GeneNet & 0.123 (0.034) & 0.000 (0.000) & 0.216 (0.053) & 0.346 (0.015) \\  \cline{2-7} 
  &  \multirow{5}{*}{$25$} & NoShrink & 0.141 (0.030) & 0.001 (0.001) & 0.234 (0.044) & 0.290 (0.016) \\ 
  &  & ShrinkNet & 0.147 (0.031) & 0.001 (0.001) & \textbf{0.242 (0.044)} & \textbf{0.307 (0.016)} \\ 
  & & SEM$_\text{L}$ & 0.063 (0.022) & 0.000 (0.000) & 0.117 (0.039) & 0.256 (0.015) \\ 
  & & GL$_{\lambda}$ & 0.038 (0.028) & 0.000 (0.000) & 0.080 (0.046) & 0.296 (0.017) \\ 
  & & GeneNet & 0.000 (0.000) & 0.000 (0.000) & 0.000 (0.000) & 0.302 (0.016) \\ 
%%%%%%%%%%%%%%%%
%%%% RANDOM
%%%%%%%%%%%%%%%%
  \hline\hline
\multirow{15}{*}{\textbf{\rotatebox{90}{RANDOM}}}& \multirow{5}{*}{$100$}& NoShrink & 0.230 (0.014) & 0.002 (0.001) & \textbf{0.367 (0.018)} & 0.317 (0.007) \\ 
  & & ShrinkNet & 0.227 (0.013) & 0.002 (0.001) & 0.365 (0.017) & \textbf{0.323 (0.006)} \\ 
  & & SEM$_\text{L}$ & 0.132 (0.012) & 0.001 (0.000) & 0.232 (0.018) & 0.309 (0.007) \\ 
  & & GL$_{\lambda}$ & 0.036 (0.023) & 0.000 (0.000) & 0.069 (0.042) & 0.315 (0.006) \\ 
  & & GeneNet & 0.080 (0.017) & 0.000 (0.000) & 0.148 (0.029) & 0.296 (0.008) \\  \cline{2-7} 
  & \multirow{5}{*}{$50$}& NoShrink & 0.143 (0.014) & 0.004 (0.001) & \textbf{0.242 (0.021)} & 0.271 (0.009) \\ 
  & & ShrinkNet & 0.136 (0.014) & 0.003 (0.001) & 0.234 (0.021) & \textbf{0.283 (0.007)} \\ 
  & & SEM$_\text{L}$ & 0.048 (0.010) & 0.000 (0.000) & 0.091 (0.018) & 0.258 (0.009) \\ 
  & & GL$_{\lambda}$ & 0.008 (0.005) & 0.000 (0.000) & 0.018 (0.010) & 0.278 (0.007) \\ 
  & & GeneNet & 0.023 (0.011) & 0.000 (0.000) & 0.045 (0.021) & 0.262 (0.009) \\  \cline{2-7} 
  & \multirow{5}{*}{$25$}& NoShrink & 0.071 (0.011) & 0.004 (0.001) & \textbf{0.127 (0.018)} & 0.225 (0.009) \\ 
  & & ShrinkNet & 0.070 (0.010) & 0.004 (0.001) & \textbf{0.127 (0.017)} & 0.233 (0.009) \\ 
  & & SEM$_\text{L}$ & 0.012 (0.005) & 0.000 (0.000) & 0.025 (0.010) & 0.208 (0.009) \\ 
  & & GL$_{\lambda}$ & 0.002 (0.002) & 0.000 (0.000) & 0.007 (0.004) & \textbf{0.234 (0.009)} \\ 
  & & GeneNet & 0.001 (0.002) & 0.000 (0.000) & 0.008 (0.005) & 0.226 (0.010) \\ 
   \hline\hline
\end{tabular}
\caption{Average and standard deviations (in parentheses) of performance measures over 100 repetitions. The best performance in average for the F-score and pAUC (partial Area Under the Curve) are boldfaced.}
\end{table}
%\vfill

All in all, the simulation study demonstrates that global shrinkage considerably improves edge ranking. For network reconstruction, the small discrepancy between ShrinkNet and NoShrink indicates that the selection procedure of Section~\ref{meth:selection} is relatively robust to edge ranking. The proposed selection procedure is also shown to outperform contenders in the most high-dimensional cases.

%%%%%%%%%%%%%%%%%%%%%%%%%%%%%%%%%%%%%%%%%%%%%%%%%%%%%%
%%%%%%%%%%%%%%%%%%%%%%%%%%%%%%%%%%%%%%%%%%%%%%%%%%%%%%
%%%	section: DATA-BASED SIMULATION
%%%%%%%%%%%%%%%%%%%%%%%%%%%%%%%%%%%%%%%%%%%%%%%%%%%%%%
%%%%%%%%%%%%%%%%%%%%%%%%%%%%%%%%%%%%%%%%%%%%%%%%%%%%%%
\section{Data-based simulation}
\label{dataSim}
In this section we employ gene expression data from The Cancer Genome Atlas (TCGA) to compare the performance of our approach in reconstructing networks with SEM$_\text{L}$,  GL$_{\lambda}$, GeneNet and NoShrink (see previous Section). Data were retrieved from the TCGA cBioPortal using the R package 'cgdsr' \citep{cgdsr2013,cerami2012}. In particular, we focus on the p53 pathway in the Breast cancer data set ($n_{\text{brca}}=526$), which comprise $p^{\text{p53}}=67$ genes, and the apoptosis pathway in the Ovarian data set ($n_{\text{ov}}=537$) that comprises $p^{\text{apopt}}=79$ genes. Since the true molecular network is not exactly known, we employ a random splitting strategy for the two data sets to assess discoveries.

\subsection{Reproducibility}
\label{dataSim:reproducibility}
To study reproducibility, we randomly split the data into a small data set where $n_{\text{small}}^{\text{p53}}\in\{134,67,34\}$ and $n_{\text{small}}^{\text{apopt}}\in\{158,79,40\}$ to achieve low-, moderate- and high-dimensional situations, and a large data set where $n_{\text{large}}^{\text{p53}}\in\{392,459,492\}$ and $n_{\text{large}}^{\text{apopt}}\in\{379,458,497\}$ (representing the complement). The large data set is then used to validate discoveries made from the small one. As a benchmark for validation we employ the edge set $\mathcal{S}_{\text{b}}$ defined by edges that are simultaneously selected by the different methods based on the large data set. Because the lack of consensus between the different methods may render $\mathcal{S}_{\text{b}}$ too small, we only compare two methods at a time.

\begin{figure}[ht]
   \begin{minipage}[c]{0.33\linewidth}
    \centering
  	$ n_{\text{small}}^{\text{apopt}}=158$
  \end{minipage}\hfill
	\begin{minipage}[c]{0.33\linewidth}
    \centering
  	$ n_{\text{small}}^{\text{apopt}}=79$
  \end{minipage}\hfill
  \begin{minipage}[c]{0.33\linewidth}
    \centering
  	$n_{\text{small}}^{\text{apopt}}=40$
  \end{minipage}\hfill
	\begin{minipage}[c]{0.33\linewidth}
    \centering
  			\includegraphics[width=1.00\textwidth]{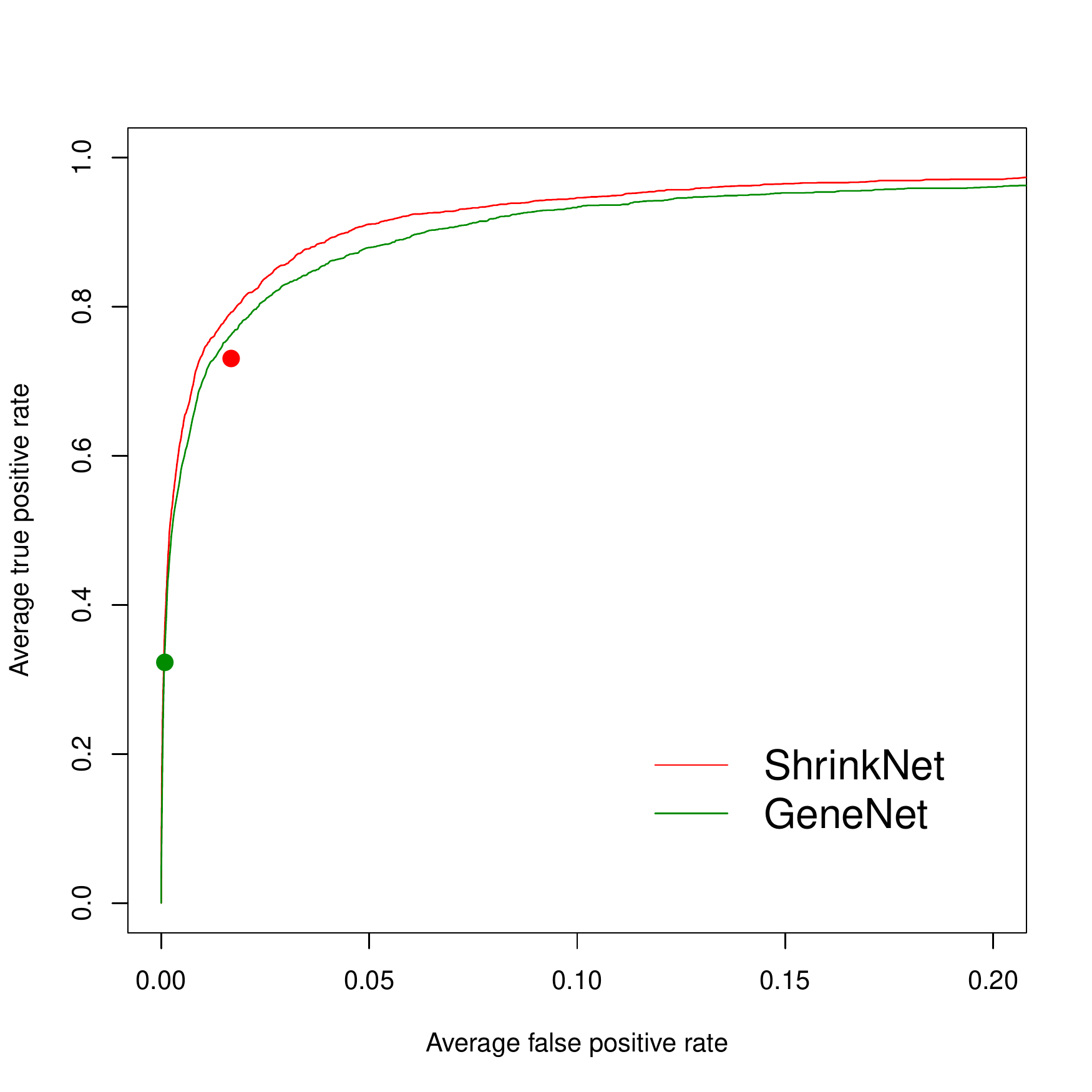}
  \end{minipage}\hfill
	\begin{minipage}[c]{0.33\linewidth}
    \centering
  			\includegraphics[width=1.00\textwidth]{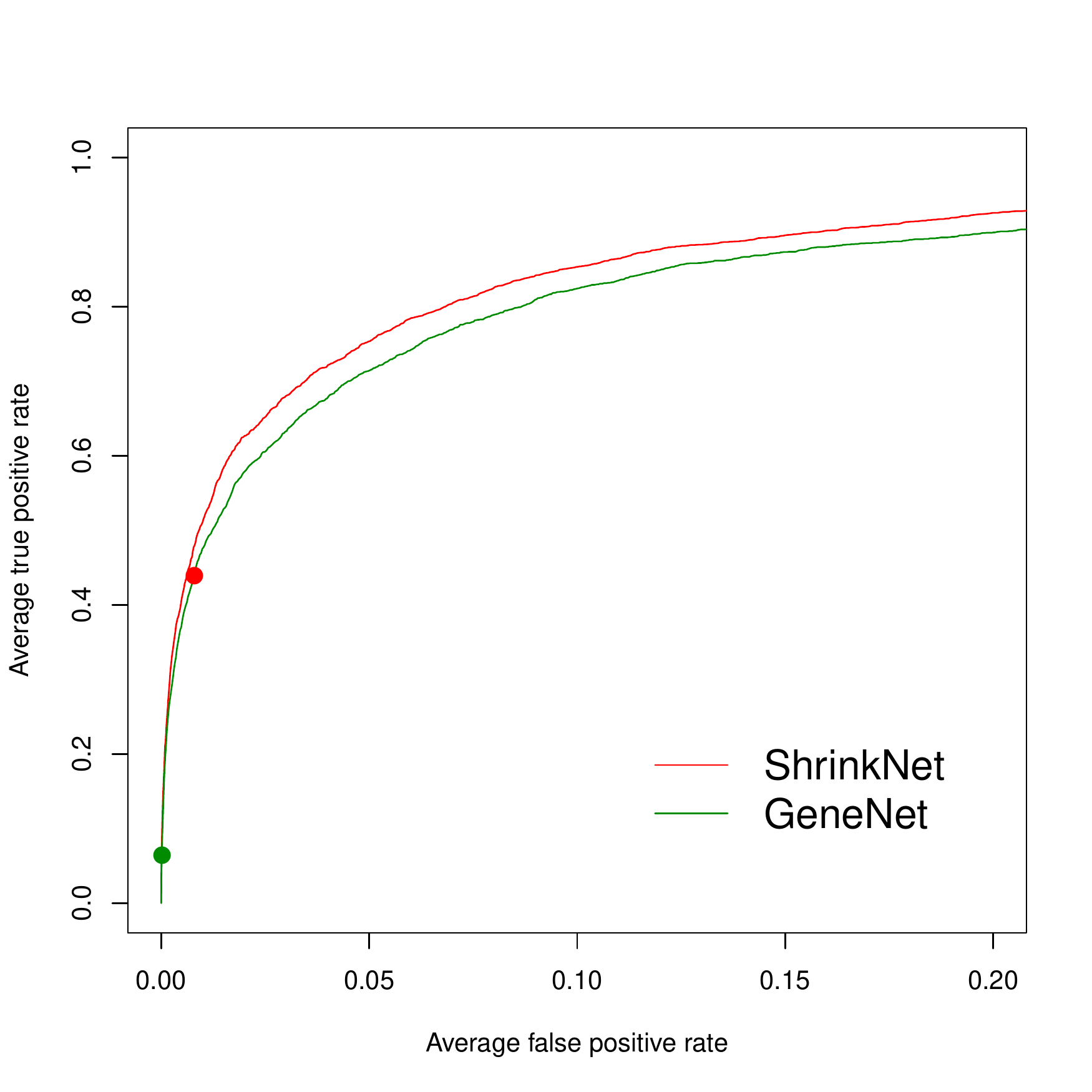}
  \end{minipage}\hfill
  \begin{minipage}[c]{0.33\linewidth}
    \centering
  			\includegraphics[width=1.00\textwidth]{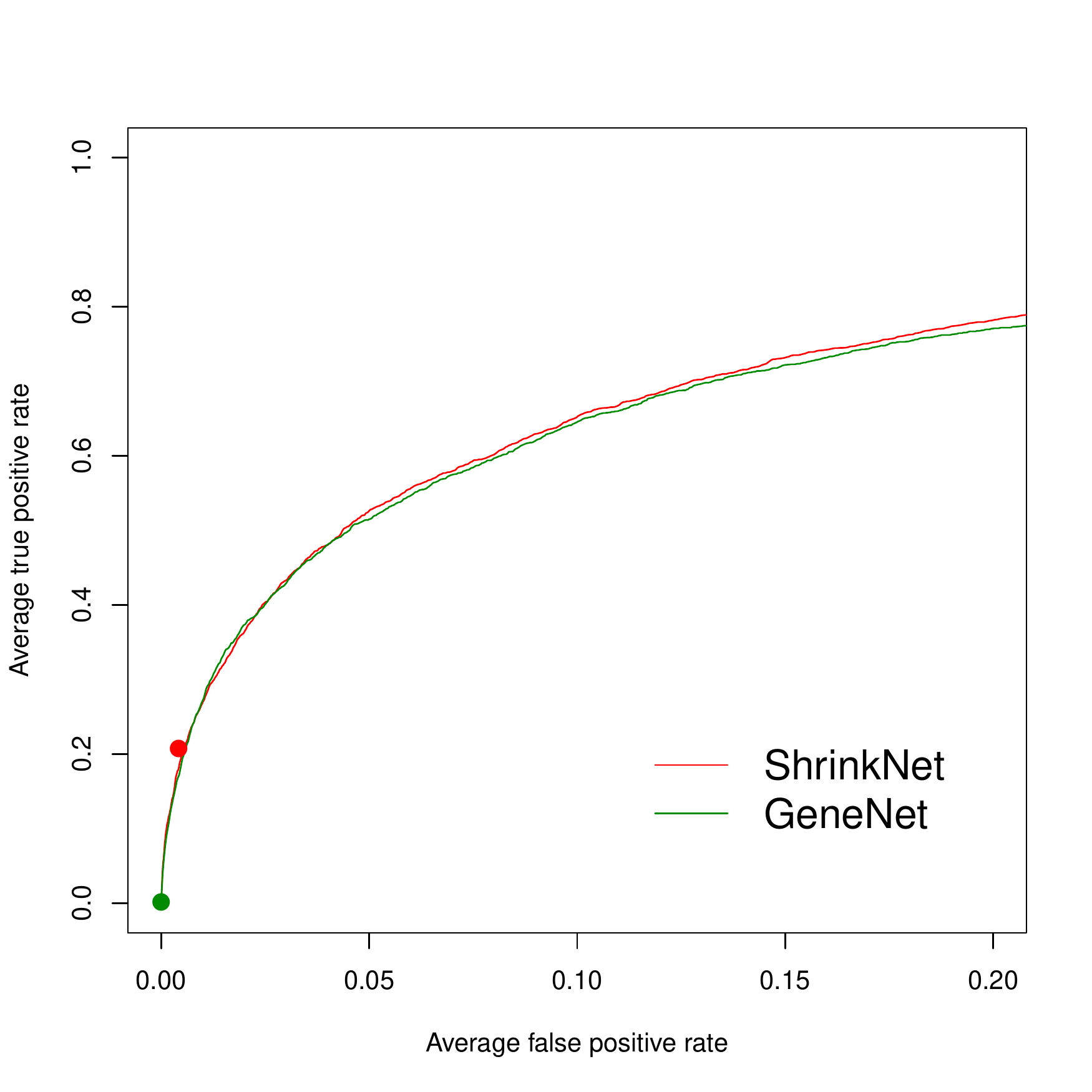}
  \end{minipage}\hfill
\vspace{-10pt}
	\begin{minipage}[c]{0.33\linewidth}
    \centering
  			\includegraphics[width=1.00\textwidth]{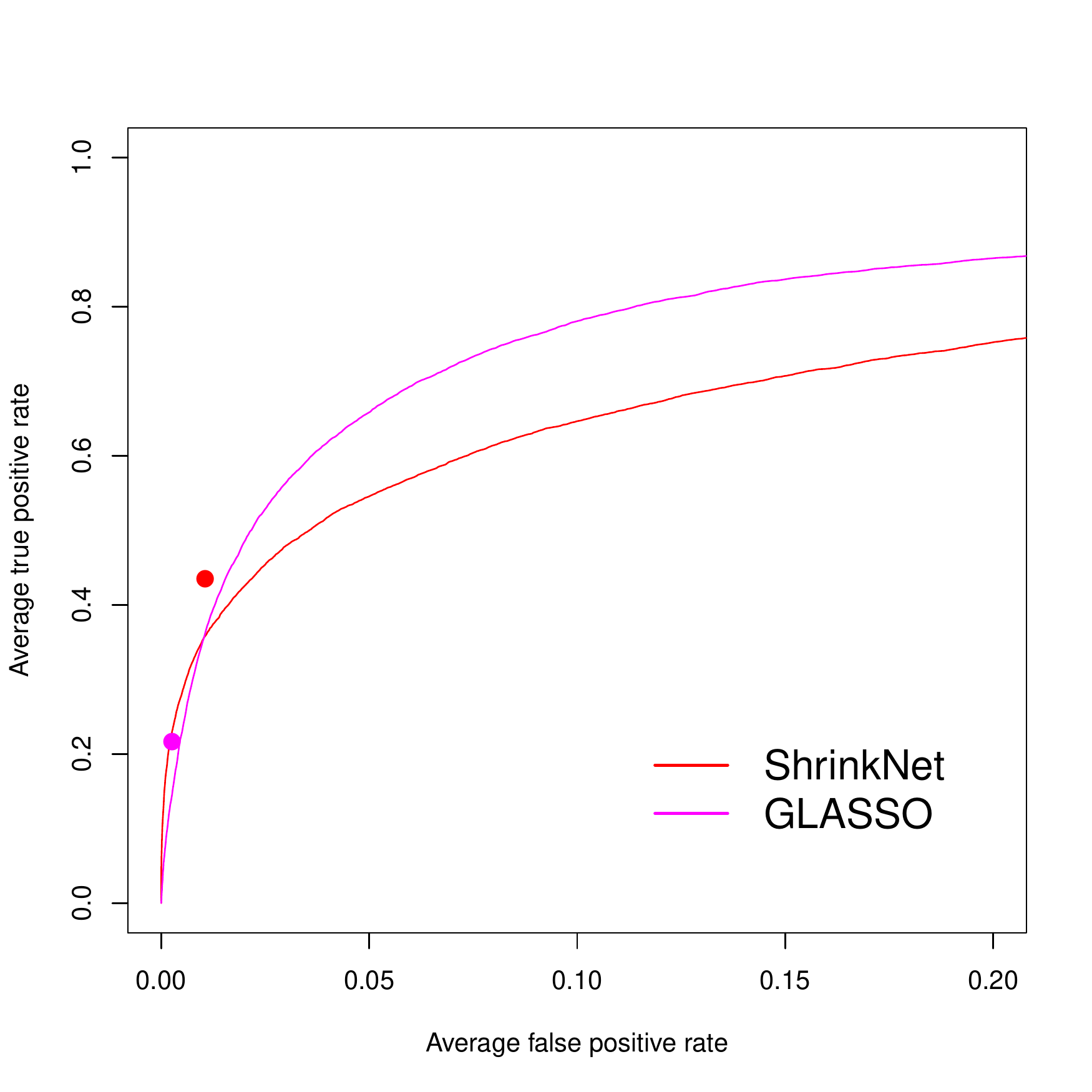}
  \end{minipage}\hfill
	\begin{minipage}[c]{0.33\linewidth}
    \centering
  			\includegraphics[width=1.00\textwidth]{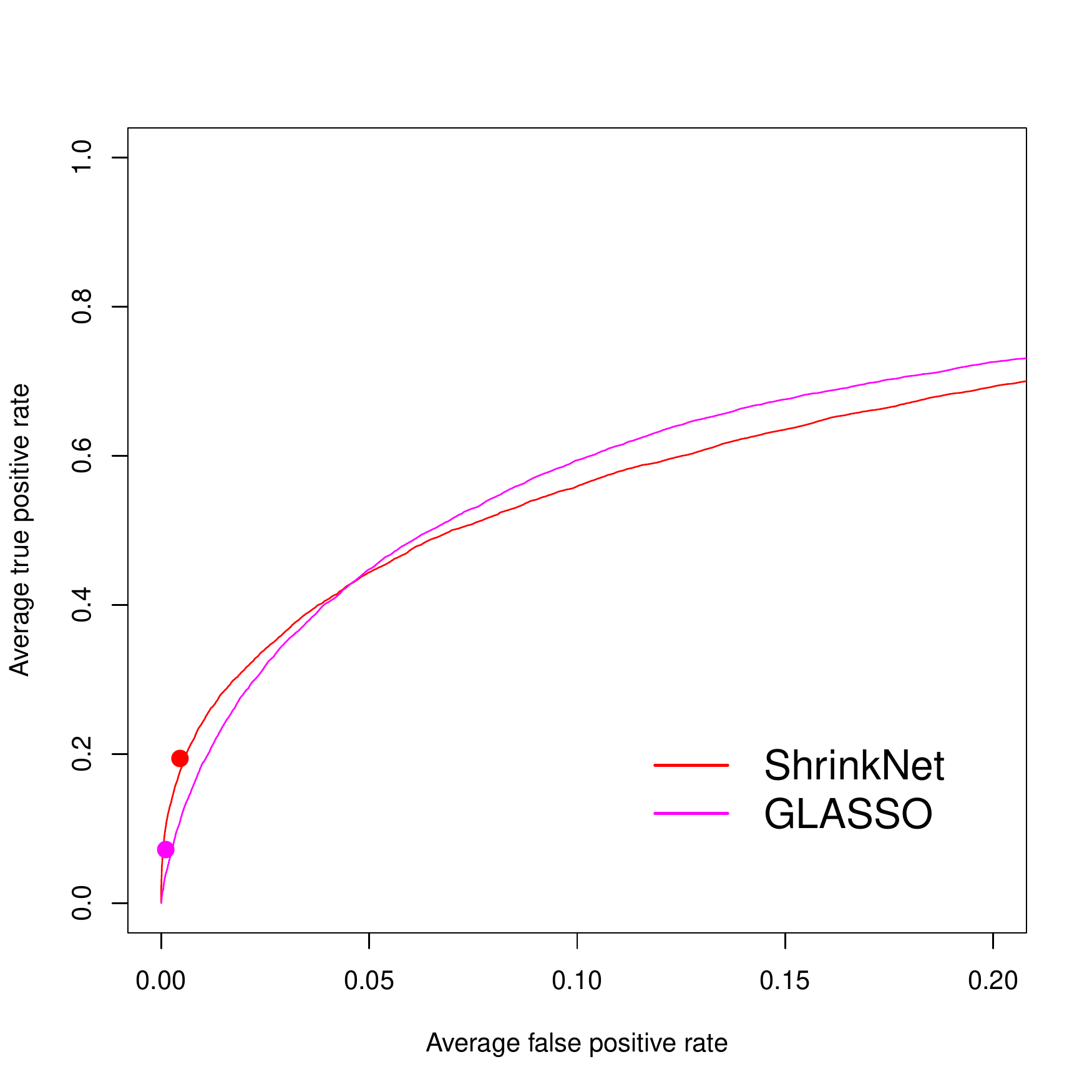}
  \end{minipage}\hfill
  \begin{minipage}[c]{0.33\linewidth}
    \centering
  			\includegraphics[width=1.00\textwidth]{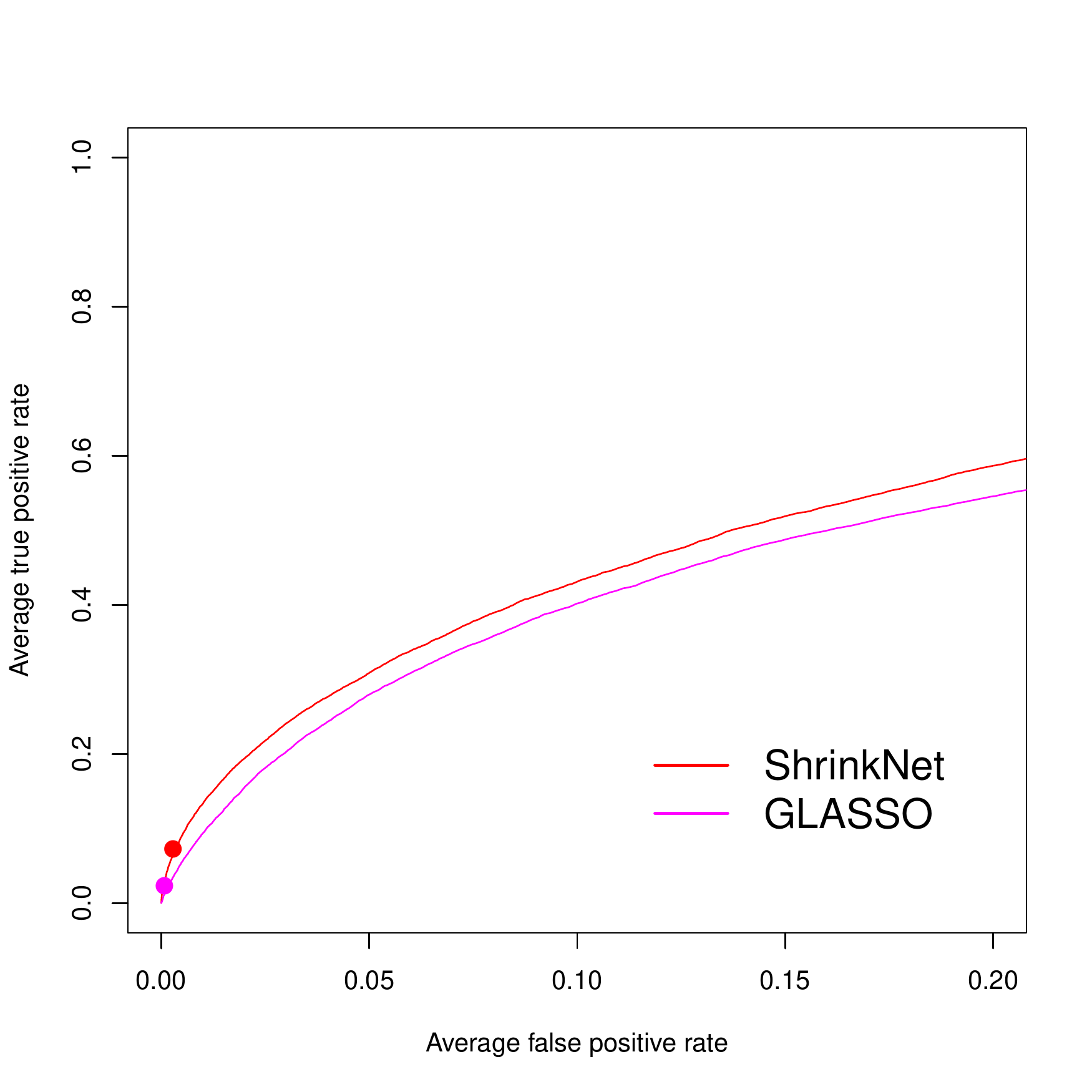}
  \end{minipage}\hfill
\vspace{-10pt}
	\begin{minipage}[c]{0.33\linewidth}
    \centering
  			\includegraphics[width=1.00\textwidth]{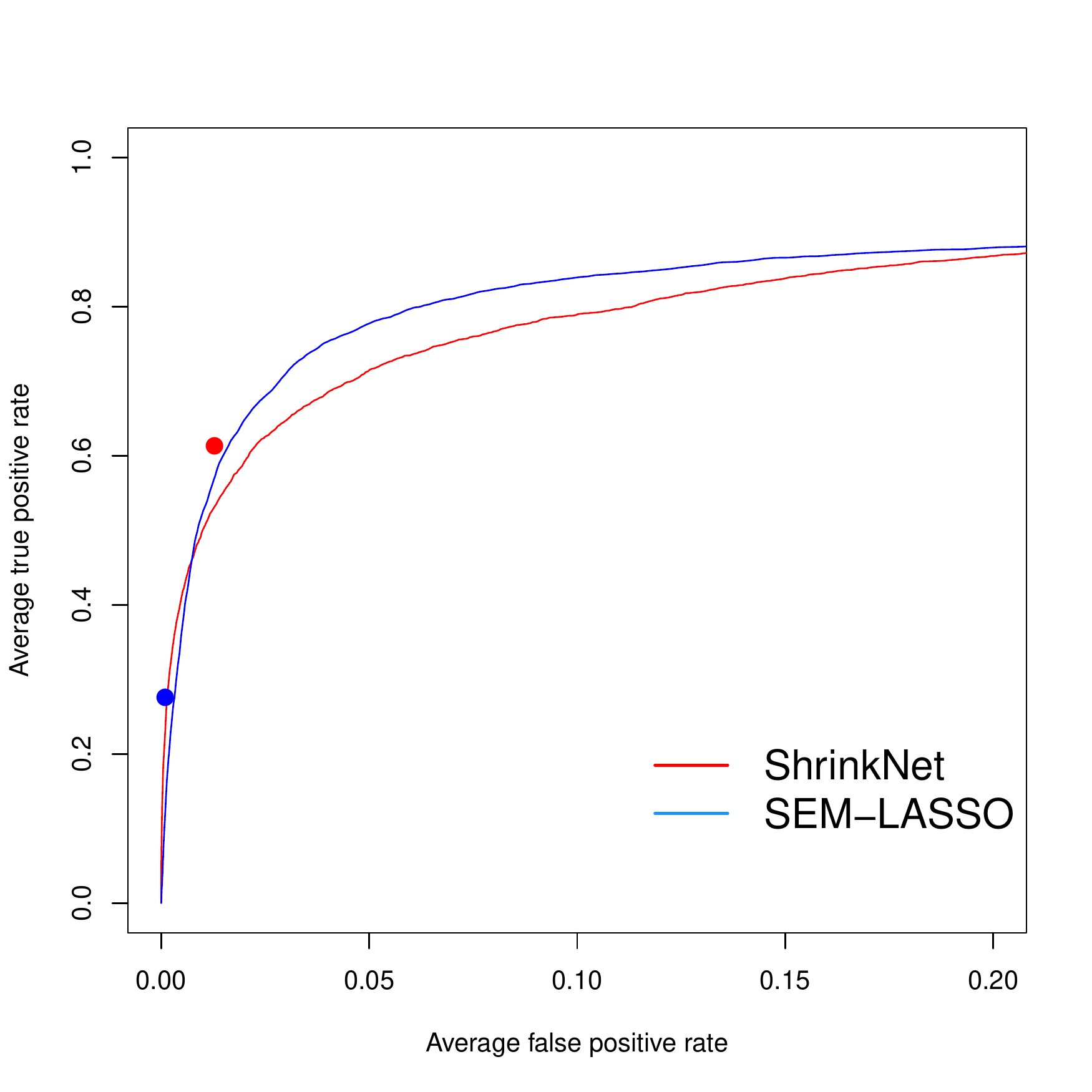}
  \end{minipage}\hfill
	\begin{minipage}[c]{0.33\linewidth}
    \centering
  			\includegraphics[width=1.00\textwidth]{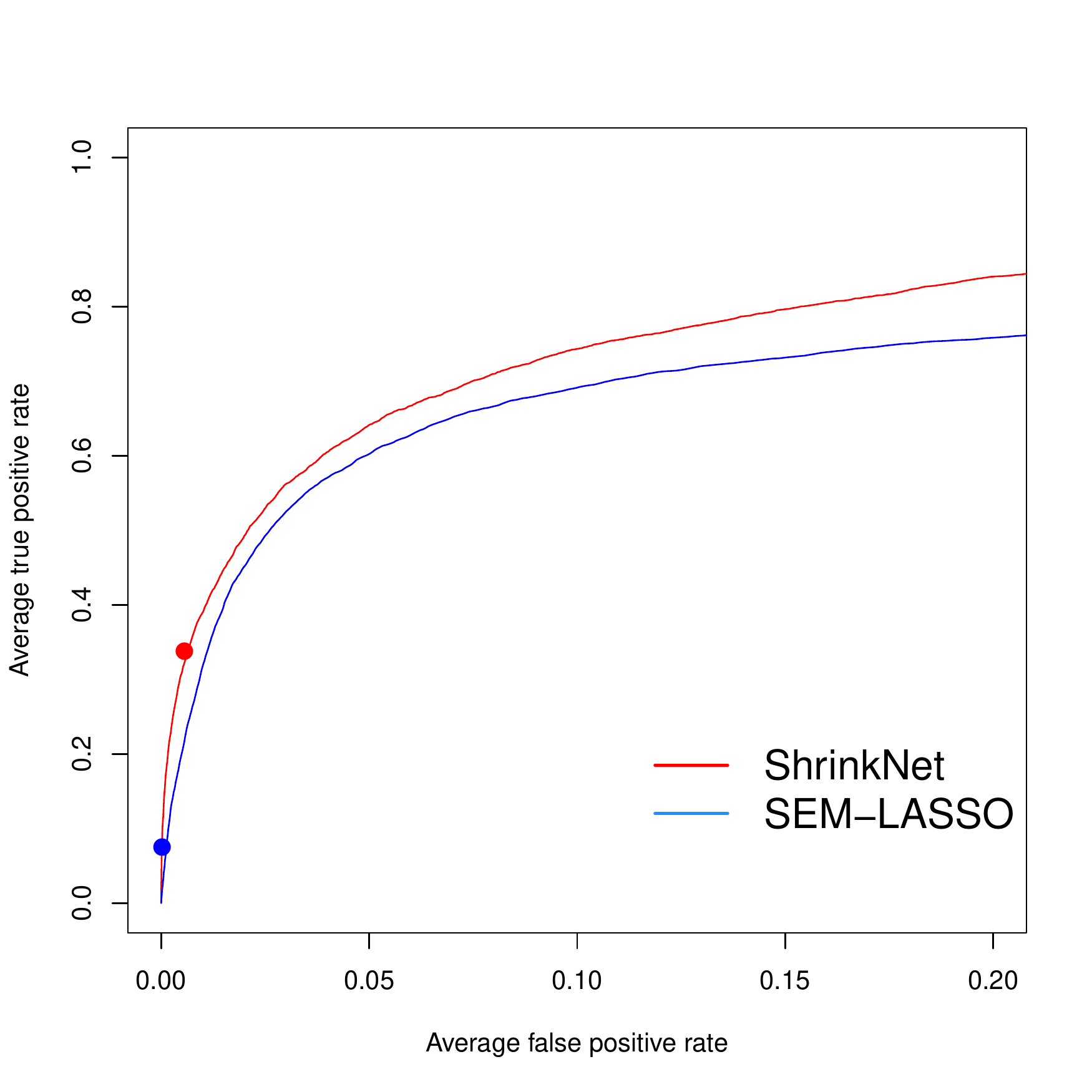}
  \end{minipage}\hfill
  \begin{minipage}[c]{0.33\linewidth}
    \centering
  			\includegraphics[width=1.00\textwidth]{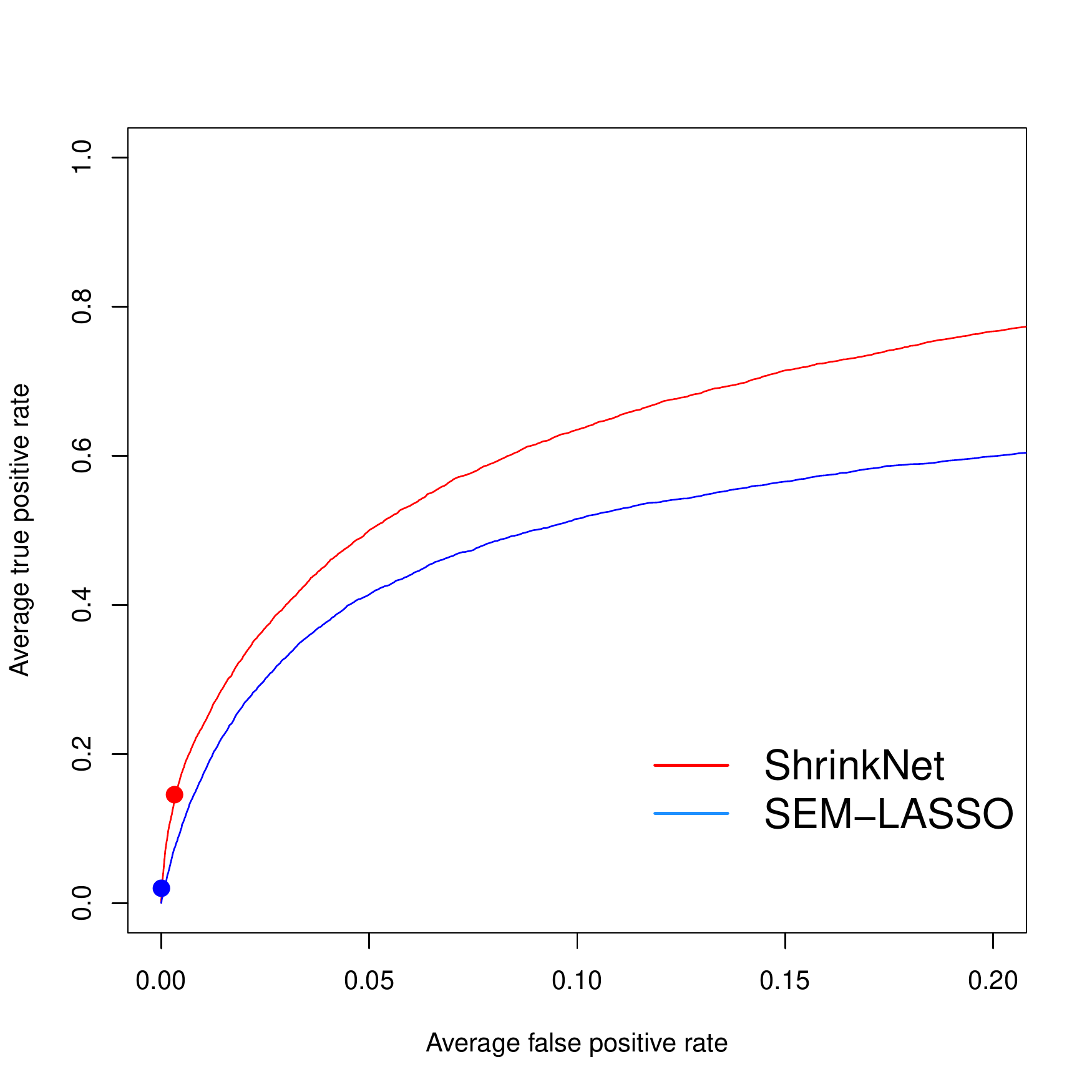}
  \end{minipage}\hfill
\vspace{-10pt}
	\begin{minipage}[c]{0.33\linewidth}
    \centering
  			\includegraphics[width=1.00\textwidth]{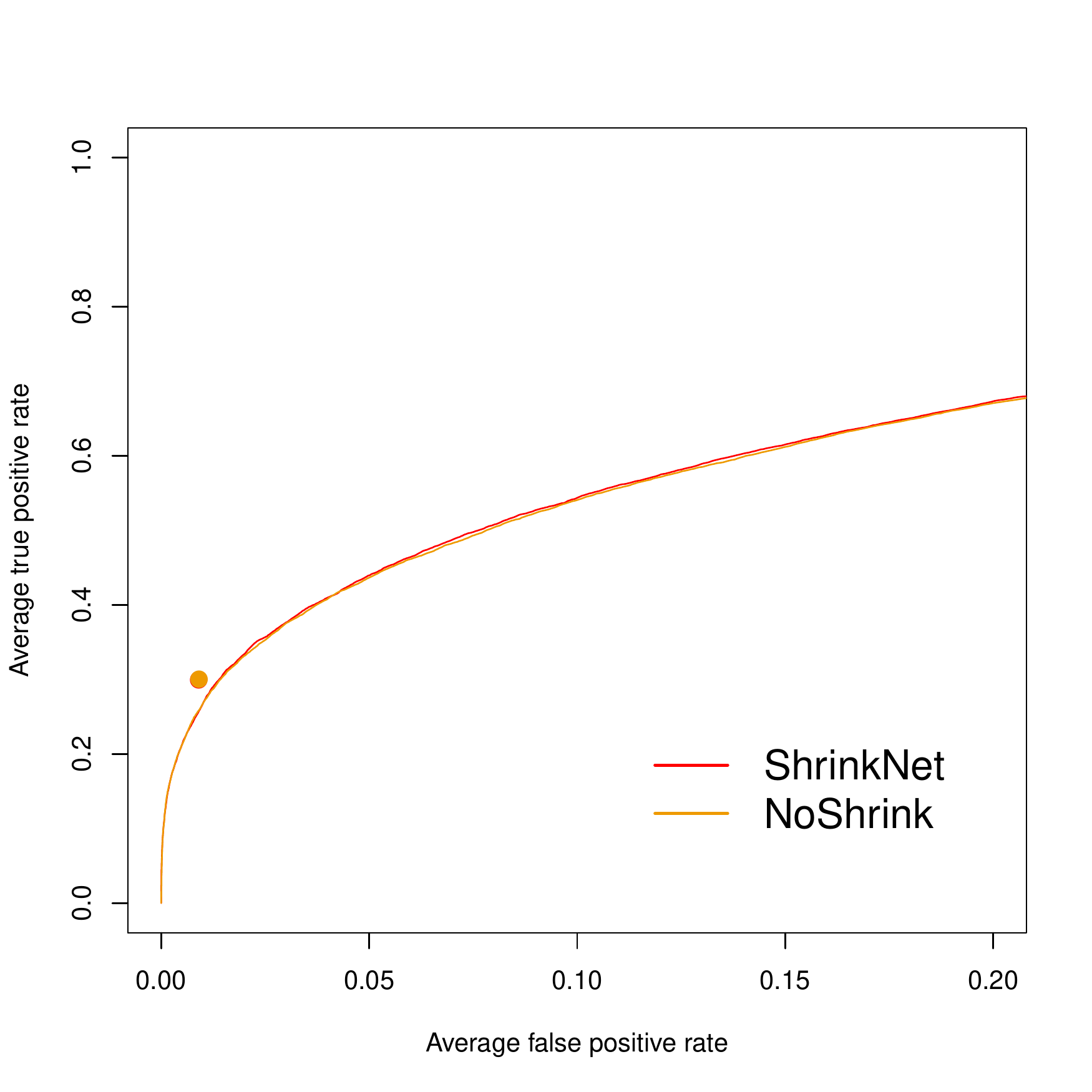}
  \end{minipage}\hfill
	\begin{minipage}[c]{0.33\linewidth}
    \centering
  			\includegraphics[width=1.00\textwidth]{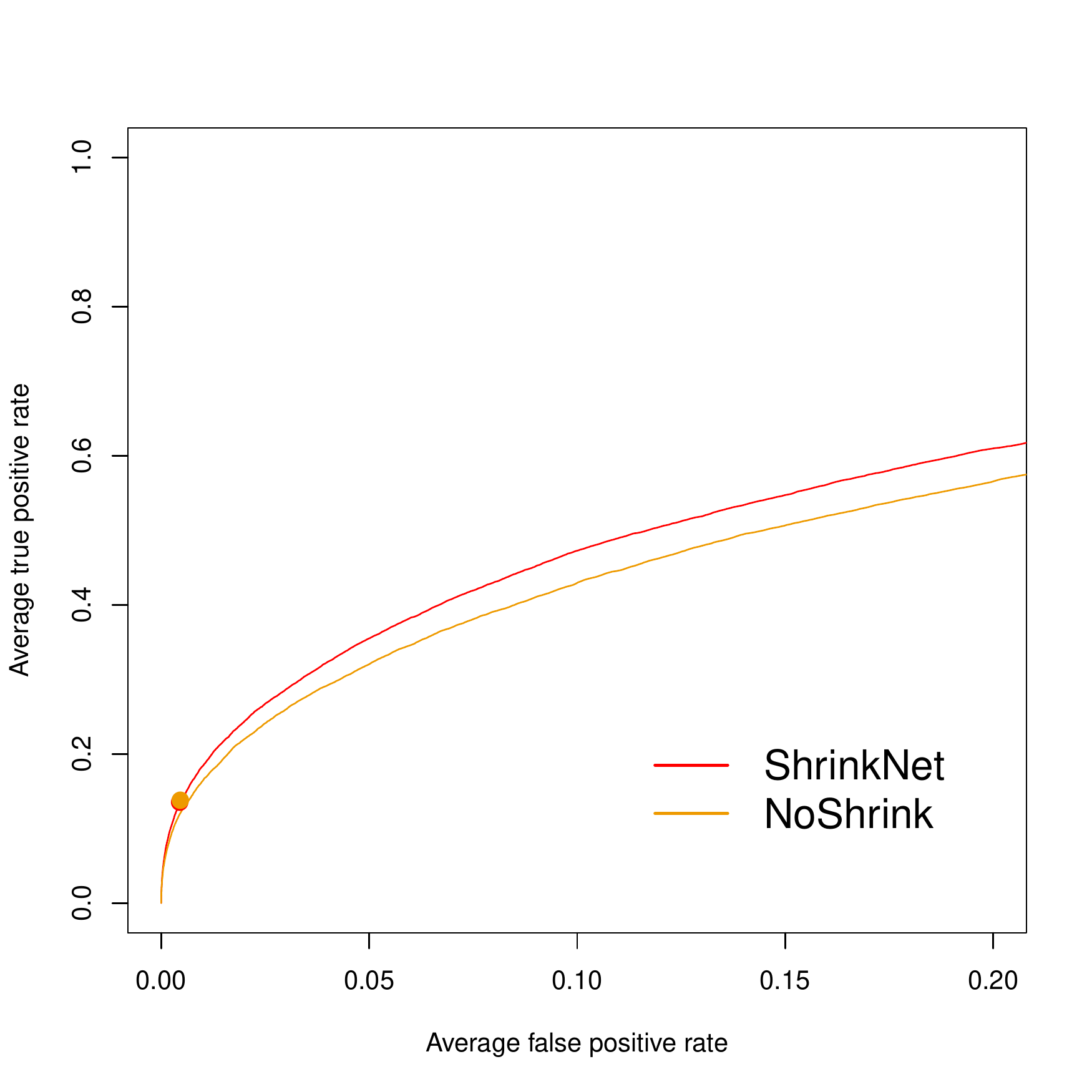}
  \end{minipage}\hfill
  \begin{minipage}[c]{0.33\linewidth}
    \centering
  			\includegraphics[width=1.00\textwidth]{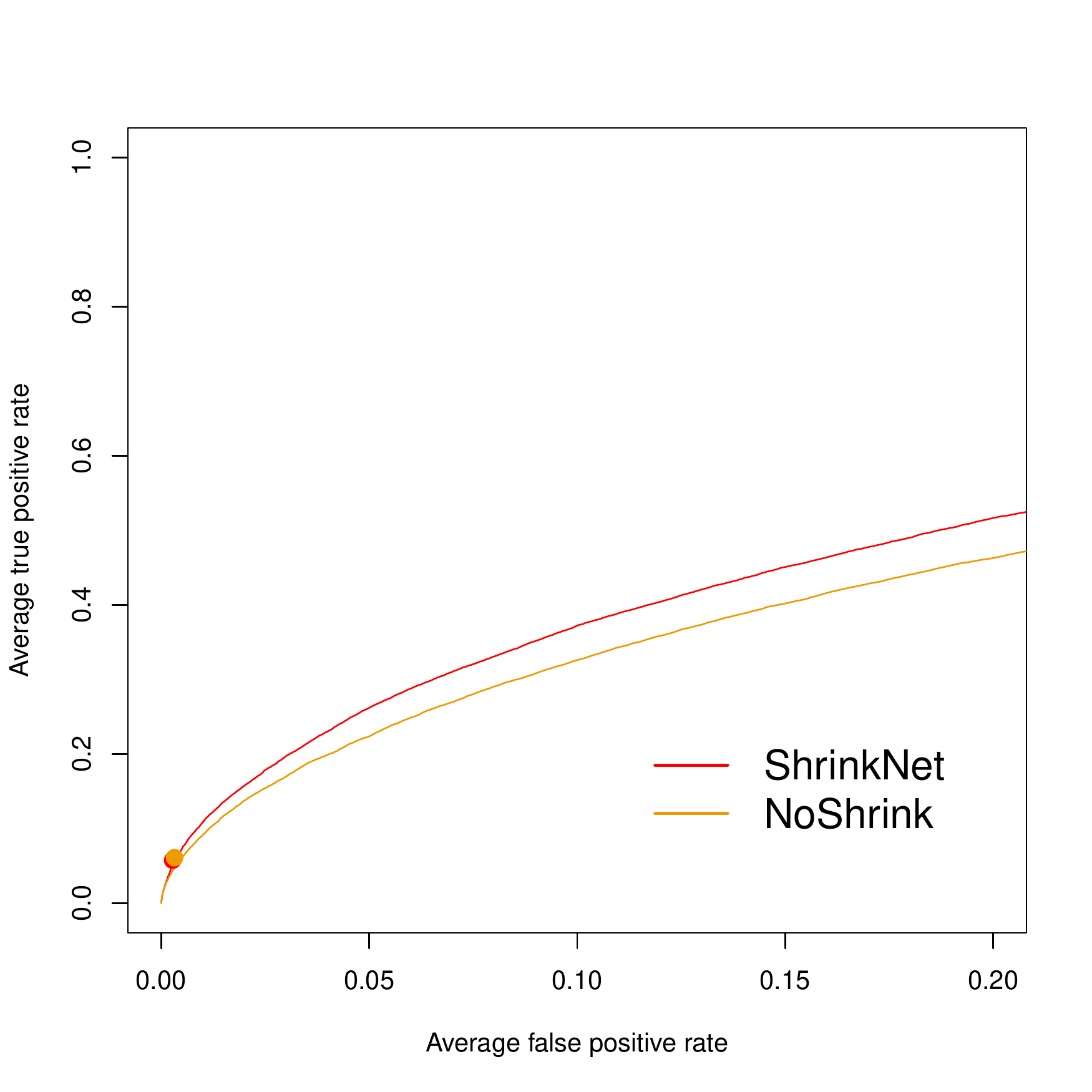}
  \end{minipage}\hfill
		\caption{Average partial ROC-curves corresponding to all pairwise comparisons of GeneNet (green), GL$_{\lambda}$ (magenta), SEM$_\text{L}$ (blue) and NoShrink (orange) with ShrinkNet (red) when the apoptosis data are randomly split into a small data set of size $n_{\text{small}}^{\text{p53}}\in\{134,67,34\}$ and a large validation one of size $n_{\text{large}}^{\text{p53}}\in\{392,459,492\}$. For each plot, dots correspond to average TPR and FPR of selected graph structures as obtained by the two inference methods under comparison.}
		\label{apoptosis}
\end{figure}

To assess performance in recovering $\mathcal{S}_{\text{b}}$ from the small data set we generate 100 random data splits and report average partial ROC curves and average TPR and FPR from the selected graphs. Figure~\ref{apoptosis} summarizes results for the four pairwise comparisons of GeneNet, SEM$_\text{L}$, GL$_{\lambda}$ and NoShrink with ShrinkNet for the apotosis pathway in the Ovarian cancer data set. Simulation results for the p53 pathway for the Breast cancer data are provided in SM Section 7. Table~\ref{dataSim:sel} and Supplementary Table~2 summarize the number of selected edges in the small and large data sets for each method.

The number of selected edges differs a lot between GeneNet, SEM$_\text{L}$, GL$_{\lambda}$ and ShrinkNet (Table \ref{dataSim:sel}). GeneNet is the most conservative approach whereas ShrinkNet selects more edges than others in the small data sets. However, when the sample size is large GL$_{\lambda}$ selects more than ShrinkNet, as illustrated by the number of discoveries in the large data sets. It is interesting to see in Table \ref{dataSim:sel} that ShrinkNet is remarkably stable in selection. The variability (as measured by the standard deviations) of the number of selected edges is relatively low, and in fact surprisingly constant in the small and large data sets, regardless of the number of selected edges. Conversely, GL$_{\lambda}$ exhibits relatively larger variability and also large differences in number of edges.\\

\begin{table}[ht]\scriptsize
\centering
\begin{tabular}{r|cc|cc|cc}
  \hline\hline
 & $n_{\text{small}}^{\text{apopt}}=$ & $n_{\text{large}}^{\text{apopt}}=$ & $n_{\text{small}}^{\text{apopt}}=$ & $n_{\text{large}}^{\text{apopt}}=$ & $n_{\text{small}}^{\text{apopt}}=$ & $n_{\text{large}}^{\text{apopt}}=$ \\ 
 & $158$ & $379$ & $79$ & $458$ & $40$ & $497$ \\  
  \hline
	ShrinkNet & 62.5 (5.7) & 138.6 (5.9) & 31.4 (5.1) & 166.9 (6) & 18.2 (4.8) & 179.6 (5.6) \\ 
	SEM$_\text{L}$ & 16.0 (3.9) & 54.0 (5.3) & 4.7 (2.3) & 65.1 (4.7) & 1.6 (1.2) & 69.2 (4) \\ 
  	GL & 25.8 (10.6) & 145.7 (35.5) & 9.6 (4.7) & 224.1 (56) & 5.3 (3.2) & 282.2 (58.1) \\ 
  	GeneNet & 10.2 (4.6) & 22.9 (4.6) & 2.2 (2.3) & 25.8 (3.5) & 0.3 (1.5) & 26.1 (2.4) \\ 
   \hline\hline
\end{tabular}
\caption{Average number of selected edges (and standard deviations in parentheses) for each method in the small and large data sets over 100 random partitioning of the Ovarian cancer data set.}
\label{dataSim:sel}
\end{table}

The results in Figure~\ref{apoptosis} suggest that ShrinkNet compares very favourably to the other methods in recovering the benchmark edge set $\mathcal{S}_{\text{b}}$. In particular, edge selection (as
represented by dots in the ROC plots) is shown to outperform the other methods clearly in all situations. In the most high-dimensional case $n_{\text{small}}^{\text{apopt}}=40$, GeneNet, SEM$_\text{L}$ and GL$_{\lambda}$ detect almost no edges in the small data set (see Table
\ref{dataSim:sel}), whereas ShrinkNet still detects a non-negligible number of edges, which translates into a higher TPR (with negligible FPR). Partial ROC curves in Figure~\ref{apoptosis} also indicate that edge ranking as provided by ShrinkNet is often superior to others. This is particularly true when
$n_{\text{small}}^{\text{apopt}}=79$ and $n_{\text{small}}^{\text{apopt}}=40$. In case $n_{\text{small}}^{\text{apopt}}=158$, SEM$_\text{L}$ and GL$_{\lambda}$ outperform ShrinkNet for edge ranking, but not for edge selection. This suggests that the selection procedure proposed in Section~\ref{meth:selection} is robust to the edge ranking on which it is based. This is confirmed by comparing ShrinkNet with NoShrink, where there is no difference in selection performance, whereas edge ranking appears to be improved by the global shrinkage prior.

\newpage

\begin{figure}[ht]
	\begin{minipage}[c]{0.33\linewidth}
    \centering
  		\subfigure[][$n_{\text{small}}^{\text{apopt}}=158$]{
  			\includegraphics[width=1.00\textwidth]{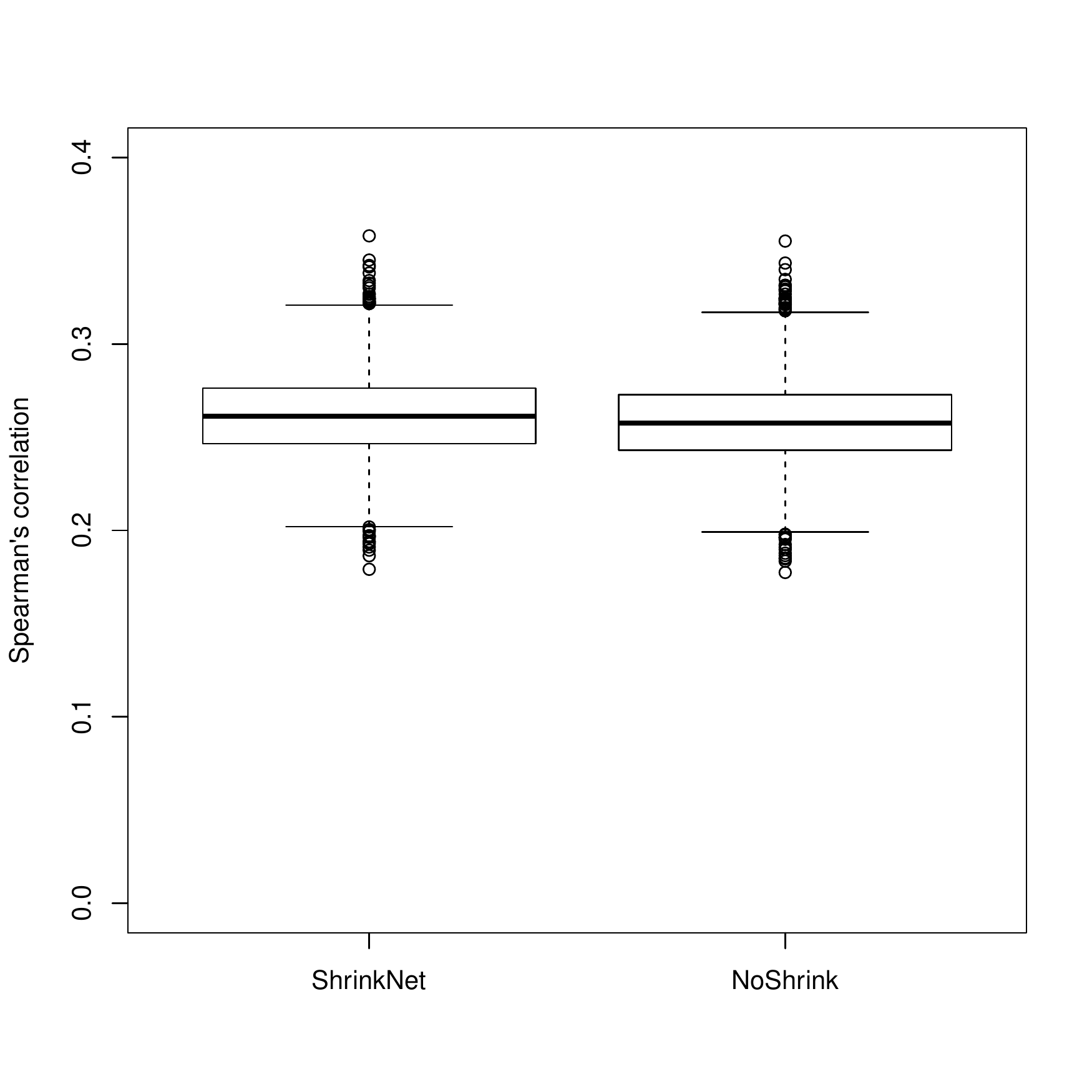}
  			\label{dataSim:stab:sub1}
  		}
  \end{minipage}\hfill
	\begin{minipage}[c]{0.33\linewidth}
    \centering
  		\subfigure[][$n_{\text{small}}^{\text{apopt}}=79$]{
  			\includegraphics[width=1.00\textwidth]{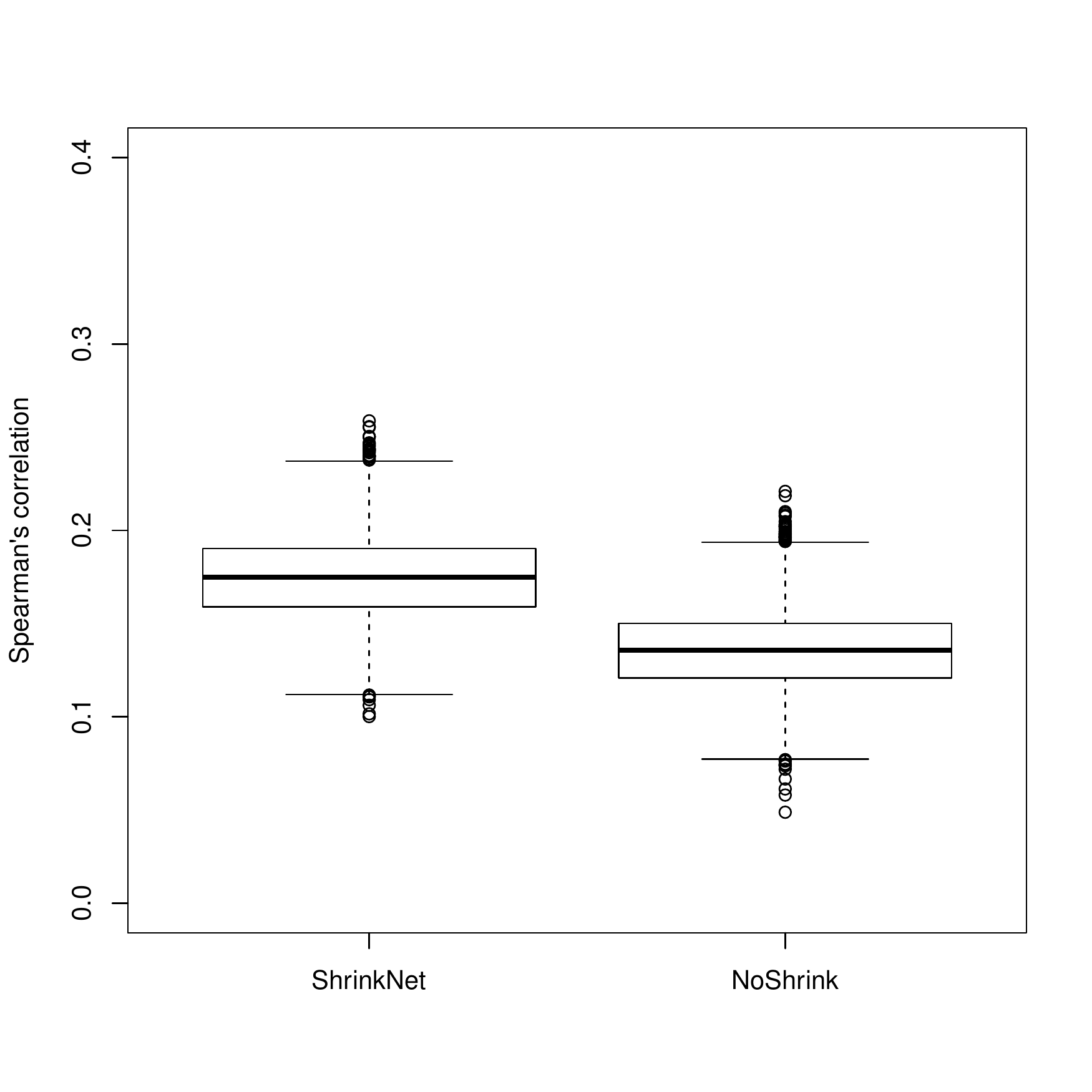}
  			\label{dataSim:stab:sub2}
  		}
  \end{minipage}\hfill
  \begin{minipage}[c]{0.33\linewidth}
    \centering
  		\subfigure[][$n_{\text{small}}^{\text{apopt}}=40$]{
  			\includegraphics[width=1.00\textwidth]{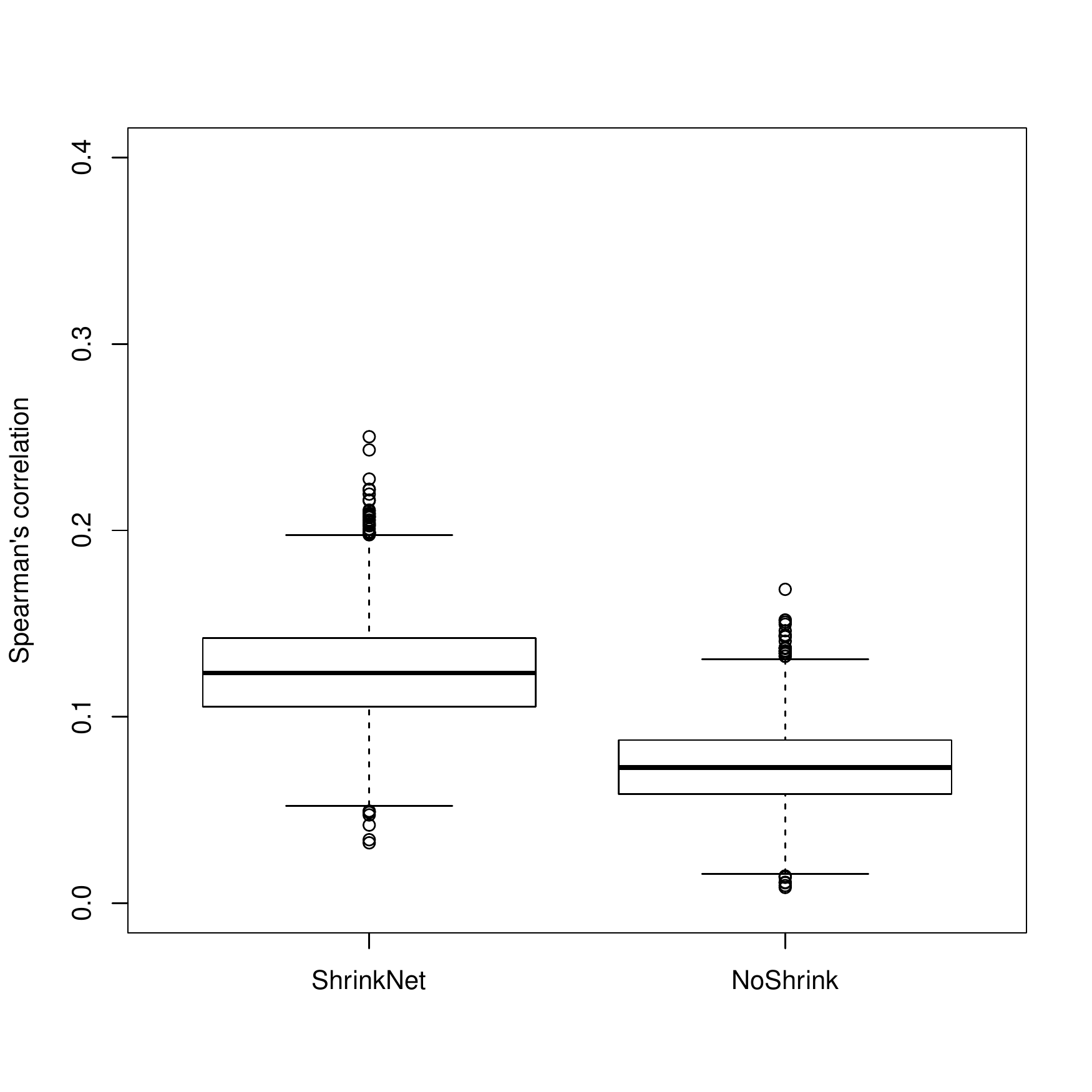}
  			\label{dataSim:stab:sub3}
  		}
  \end{minipage}\hfill
		\caption{Correlations of edge ranking as provided by ShinkNet and NoShrink across the 100 generated small data sets of size $n_{\text{small}}^{\text{apopt}}$. Each boxplot displays Spearman rank correlations between the values of $\bar{\kappa}_r$,  $r=1,\ldots,P$, obtained from all the $(100\times 99)/2=4950$ pairs of data sets of size $n_{\text{small}}^{\text{apopt}}$ for each of the two methods. Note that one does not expect high rank correlation when considering all edges.}
		\label{dataSim:stab}
\end{figure}

Finally Figure~\ref{dataSim:stab} displays rank correlation of edges between all pairs of data sets of size $n_{\text{small}}^{\text{apopt}}$ for ShrinkNet and NoShrink. The correlations are clearly higher for ShrinkNet than for NoShrink when $n_{\text{small}}^{\text{apopt}}\in\{79,40\}$, which indicates that the global shrinkage improves the stability and, hence, reproducibility of edge ranking when the sample size $n_{\text{small}}^{\text{apopt}}$ is not large.

In summary, the simulation study demonstrates that ShrinkNet reproduces better than SEM$_\text{L}$, GL$_{\lambda}$ and GeneNet on real data.\\

\subsection{Stability}
\label{dataSim:stability}
In this section, the random splitting strategy is used to study the stability of edges selected by each method. Let $\hat{\pi}_{ij}$ be the empirical selection probability of edge $(i,j)$ for a given method over the 100 generated small data sets of size $n_{\text{small}}^{\text{apopt}}$. We define the set of stable edges by $S_{\text{stable}}=\{ (i,j): \hat{\pi}_{ij}\geq \pi_{\text{thr}}\}$ where $0.5< \pi_{\text{thr}}\leq 1$. To determine an appropriate cut-off $\pi_{\text{thr}}$, which is comparable between methods, we use the stability criterion proposed by \citep{meinshausen2010}. This is based on the following upper bound on the expected number $\mathbb{E}(V)$ of falsely selected edges:
\begin{equation}
	\label{SS}
	\mathbb{E}(V)\leq \frac{q^2}{(2\pi_{\text{thr}}-1)P},
\end{equation}
where $q$ is the expected number of edges selected by the given method and $P$ is the total number of edges ($P_{\text{apopt}}=3081$ and $P_{\text{p53}}=2211$). To compare the set of stable edges between the different methods, we set $\mathbb{E}(V)=30$ as in \citet{meinshausen2010}. Then, $\pi_{\text{thr}}$ (and hence $S_{\text{stable}}$) is determined using an empirical estimate of $q$ (see Table \ref{dataSim:sel} and SM Table 2). Because the type I error is controlled in the same way for all methods, comparison can reasonably be based on the number of stable edges.

\begin{figure}[ht]
   \begin{minipage}[c]{0.33\linewidth}
    \centering
  	$n_{\text{small}}^{\text{apopt}}=158$
  \end{minipage}\hfill
	\begin{minipage}[c]{0.33\linewidth}
    \centering
  	$n_{\text{small}}^{\text{apopt}}=79$
  \end{minipage}\hfill
  \begin{minipage}[c]{0.33\linewidth}
    \centering
  	$n_{\text{small}}^{\text{apopt}}=40$
  \end{minipage}\hfill
	\begin{minipage}[c]{0.33\linewidth}
    \centering
  			\includegraphics[trim=90 85 65 70,clip,width=1.00\textwidth]{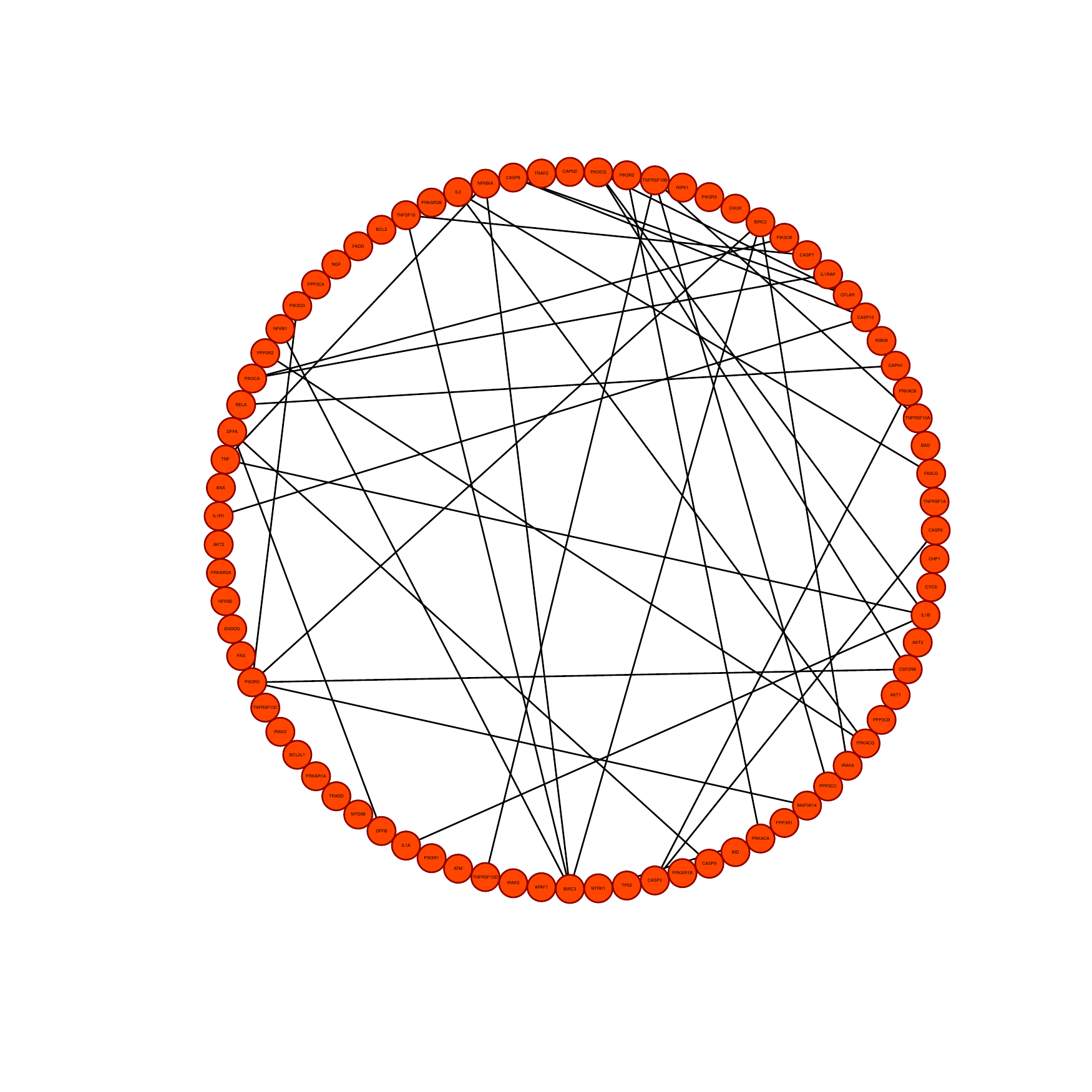}
  \end{minipage}\hfill
	\begin{minipage}[c]{0.33\linewidth}
    \centering
  			\includegraphics[trim=90 85 65 70,clip,width=1.00\textwidth]{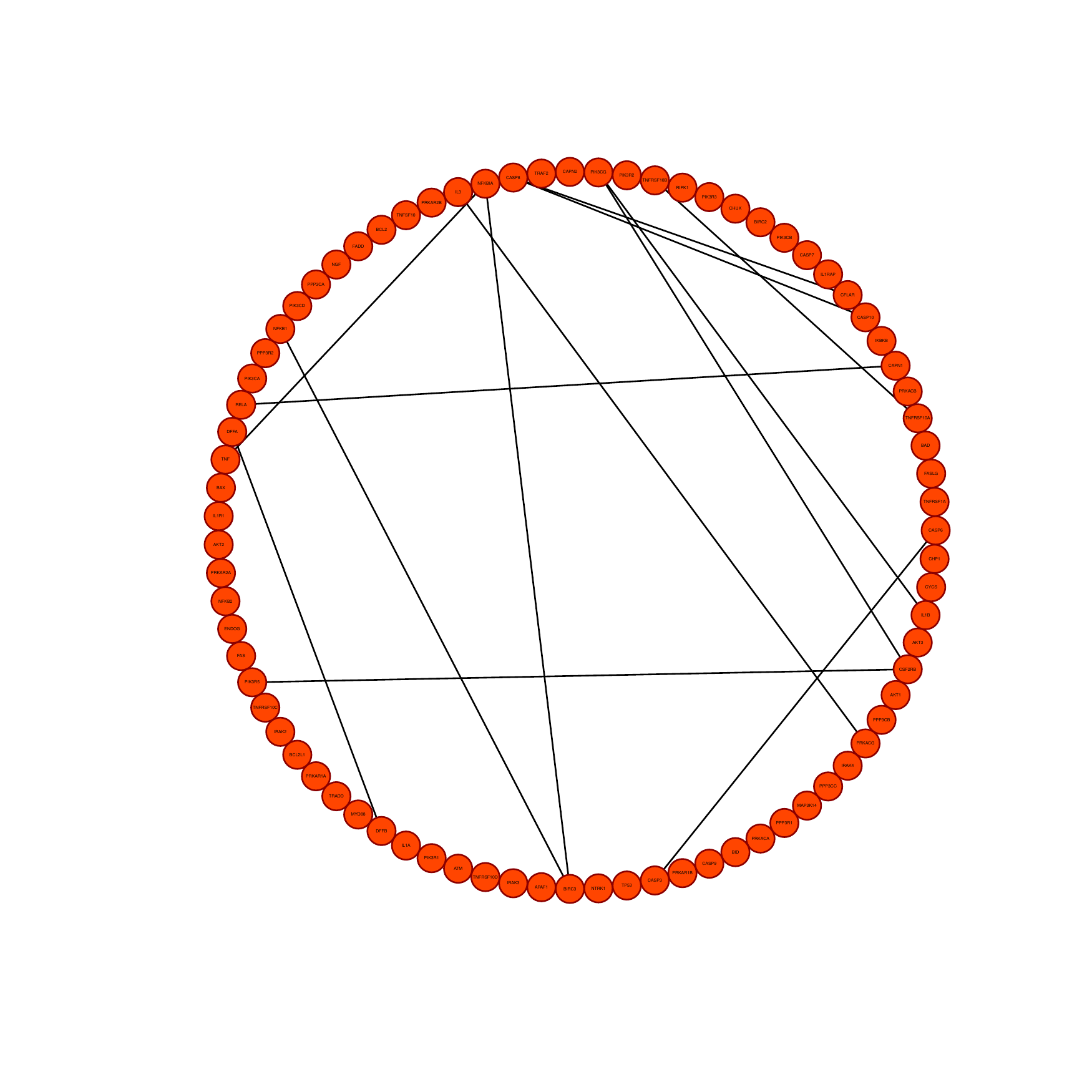}
  \end{minipage}\hfill
  \begin{minipage}[c]{0.33\linewidth}
    \centering
  			\includegraphics[trim=90 85 65 70,clip,width=1.00\textwidth]{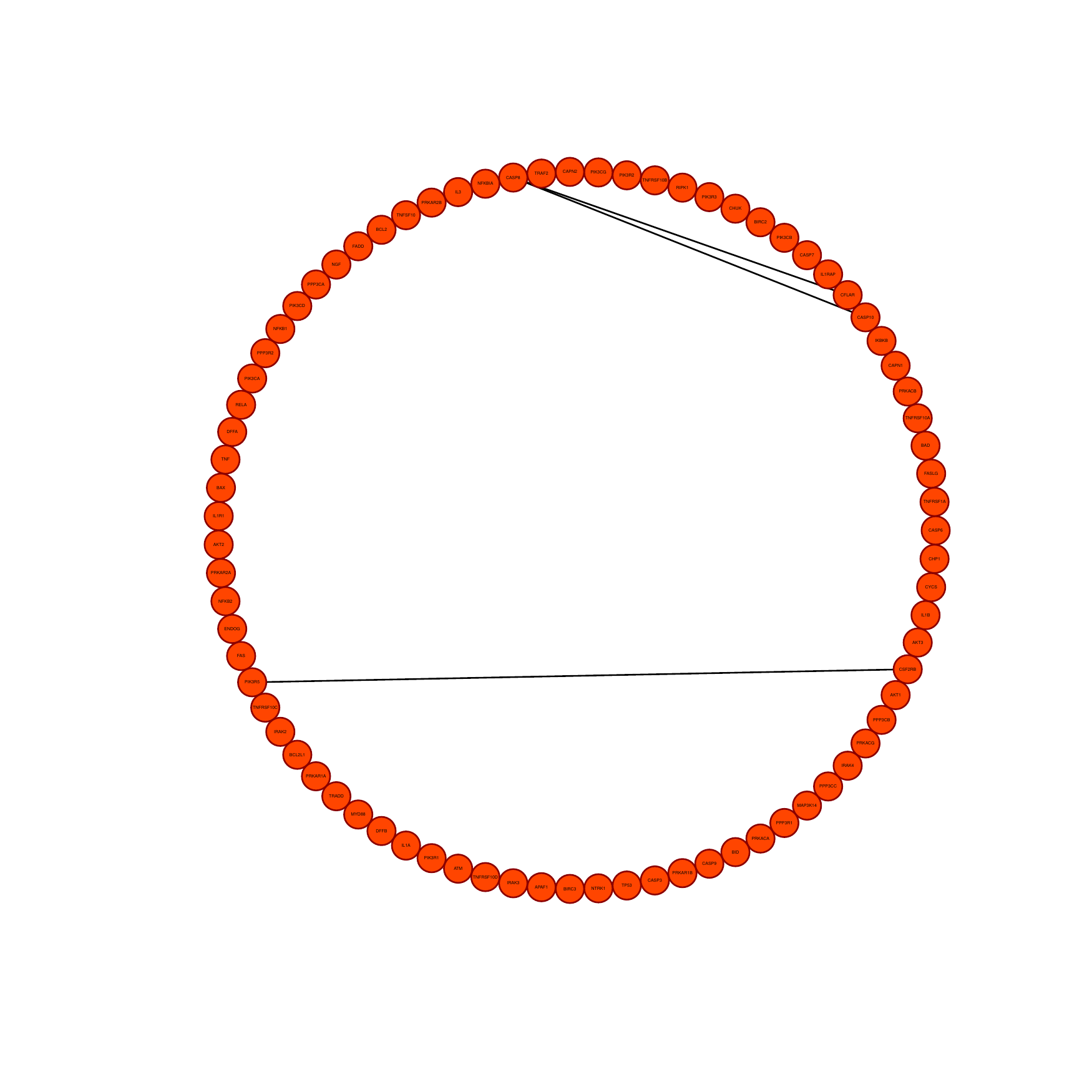}
  \end{minipage}\hfill
  \begin{minipage}[c]{0.33\linewidth}
  			\includegraphics[trim=90 85 65 70,clip,width=1.00\textwidth]{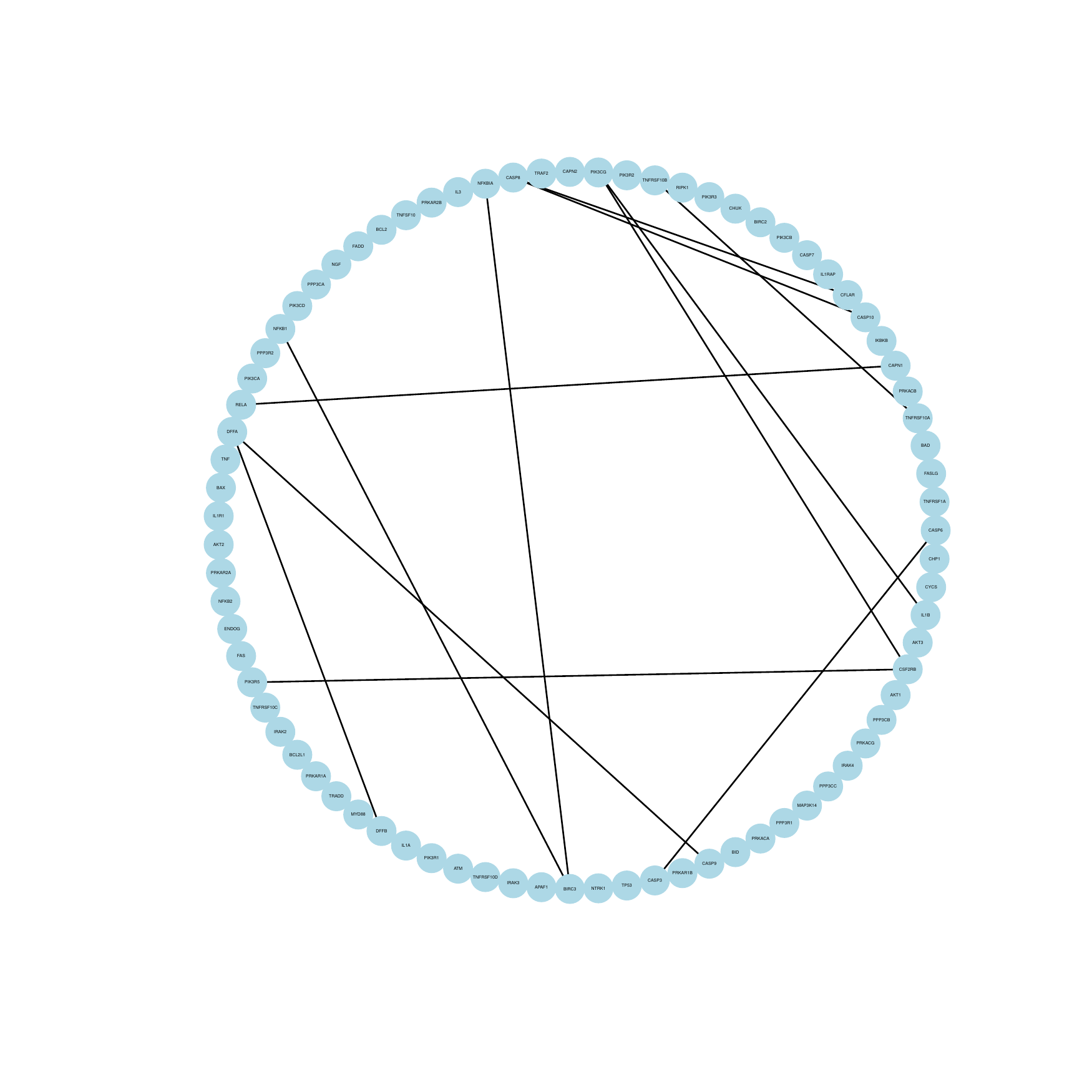}
  \end{minipage}\hfill
	\begin{minipage}[c]{0.33\linewidth}
    \centering
 			\includegraphics[trim=90 85 65 70,clip,width=1.00\textwidth]{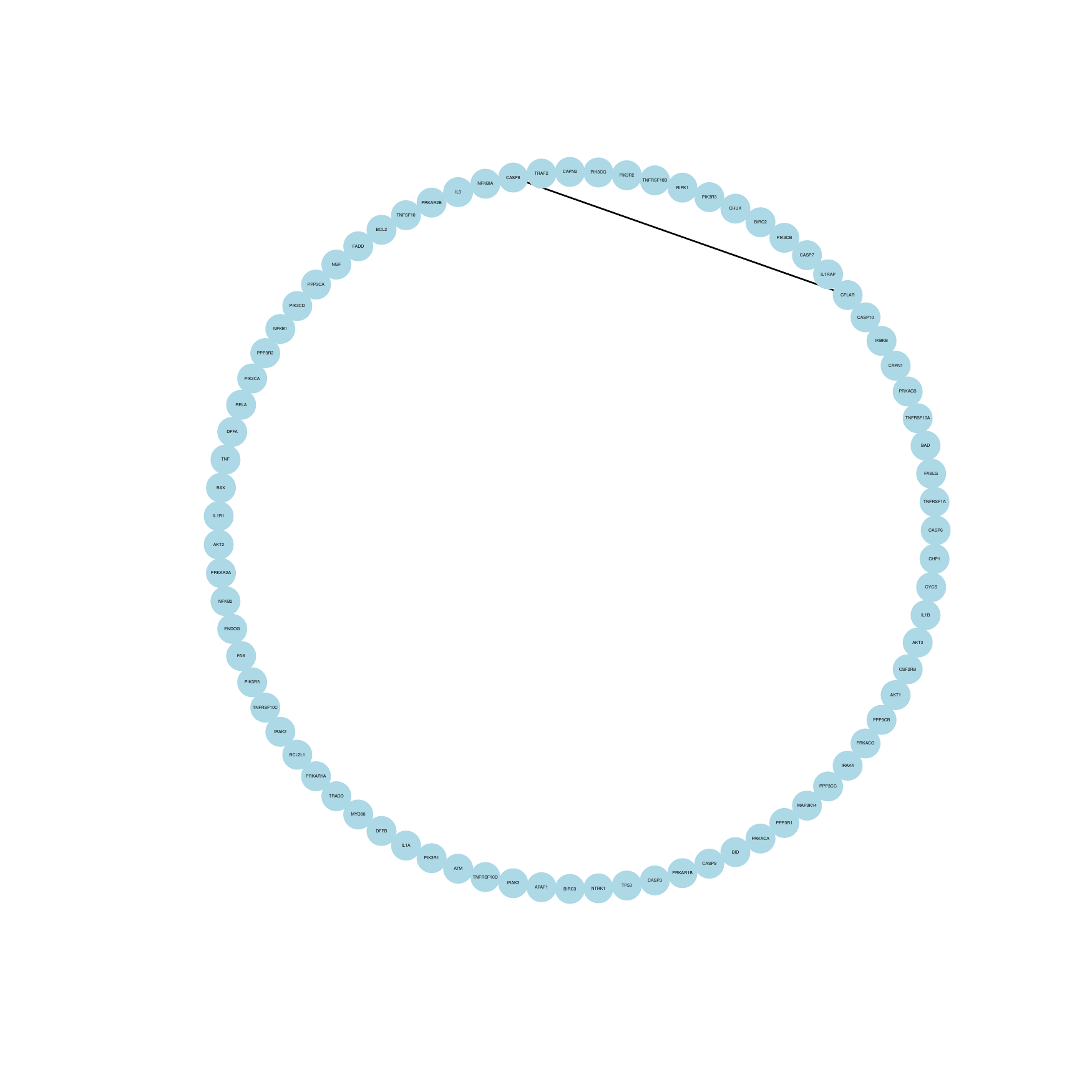}
  \end{minipage}\hfill
  \begin{minipage}[c]{0.33\linewidth}
    \centering
  			\includegraphics[trim=90 85 65 70,clip,width=1.00\textwidth]{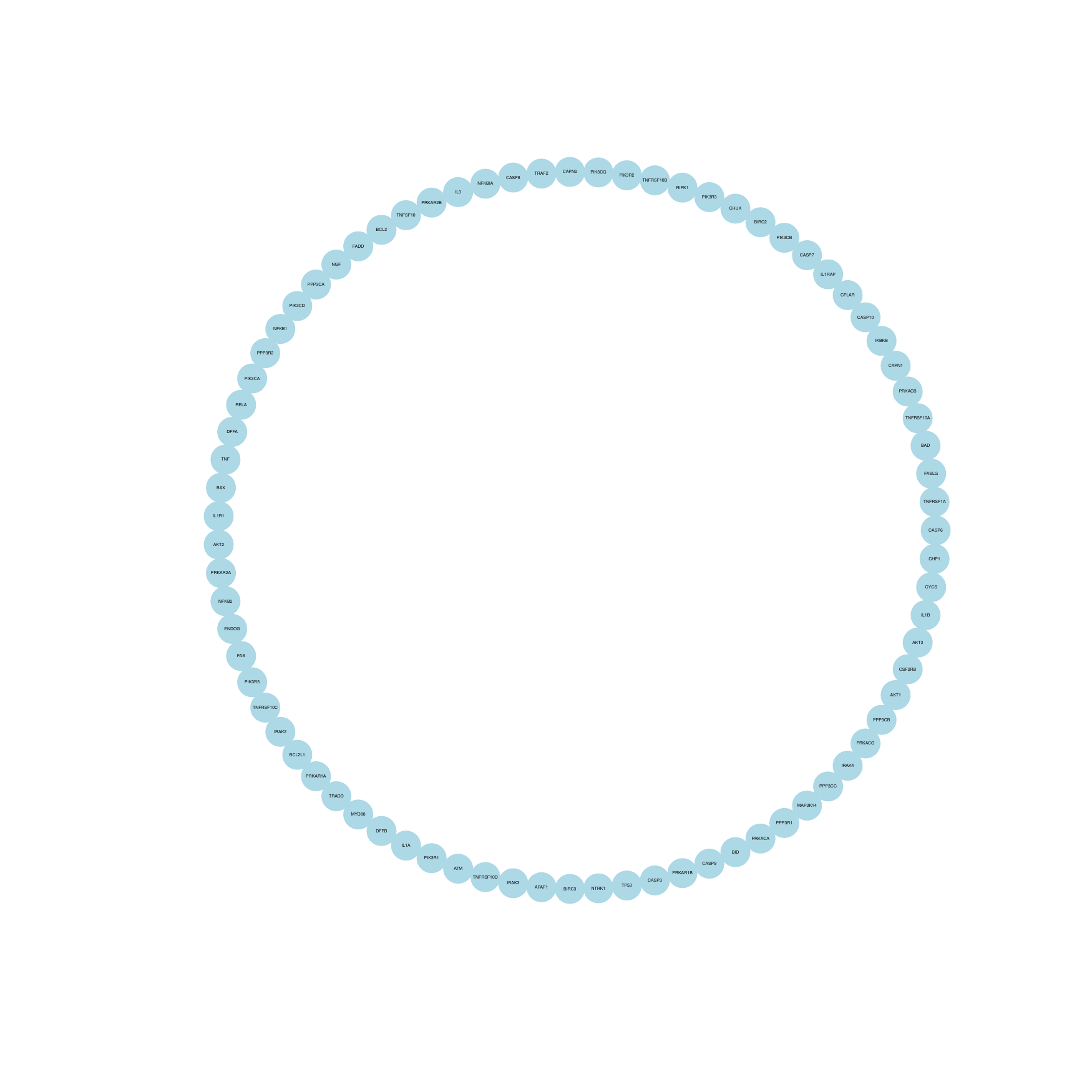}
  \end{minipage}\hfill
 \begin{minipage}[c]{0.33\linewidth}
    \centering
  			\includegraphics[trim=90 85 65 70,clip,width=1.00\textwidth]{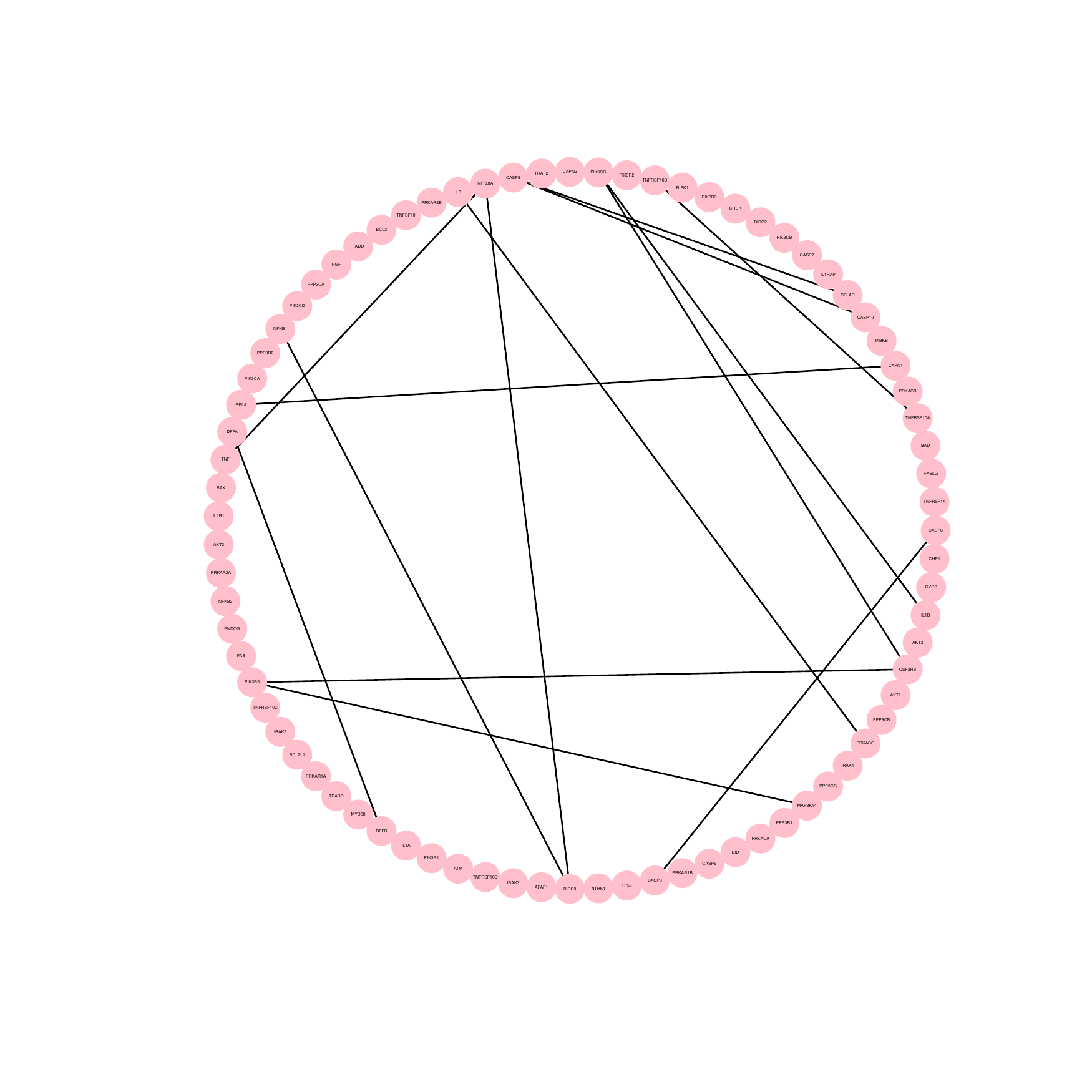}
  \end{minipage}\hfill
	\begin{minipage}[c]{0.33\linewidth}
    \centering
  			\includegraphics[trim=90 85 65 70,clip,width=1.00\textwidth]{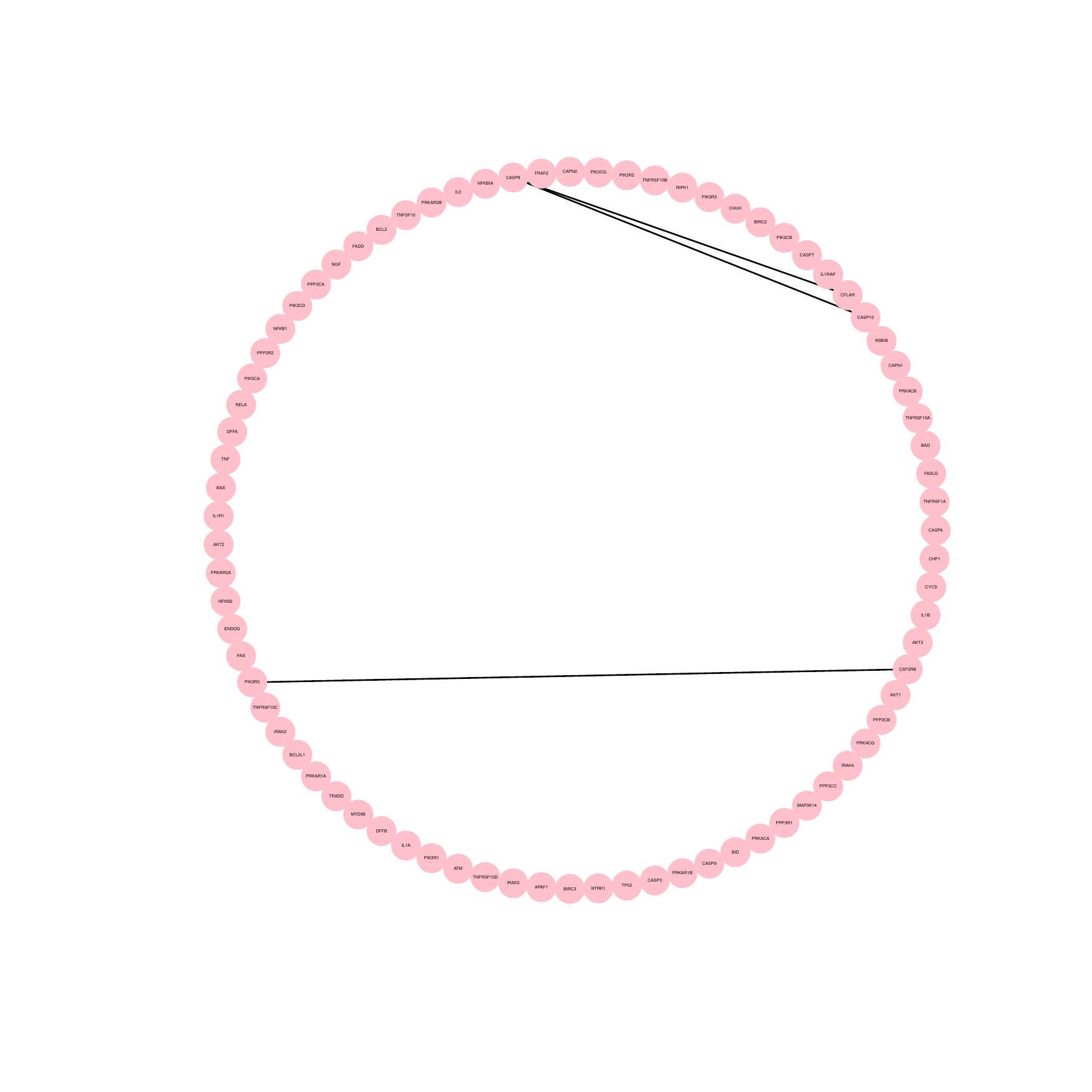}
  \end{minipage}\hfill
  \begin{minipage}[c]{0.33\linewidth}
    \centering
  			\includegraphics[trim=90 85 65 70,clip,width=1.00\textwidth]{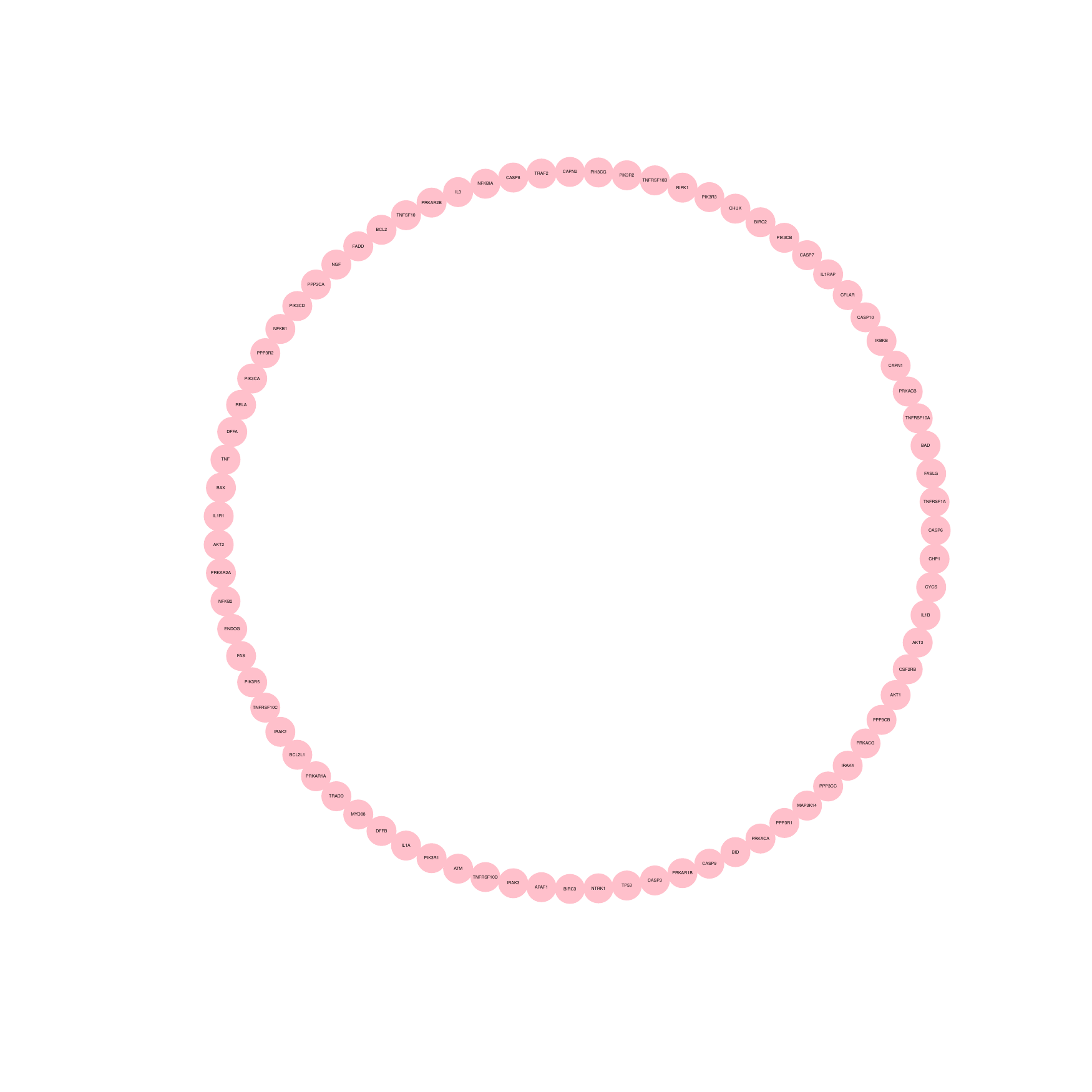}
  \end{minipage}\hfill
 \begin{minipage}[c]{0.33\linewidth}
    \centering
  			\includegraphics[trim=90 85 65 70,clip,width=1.00\textwidth]{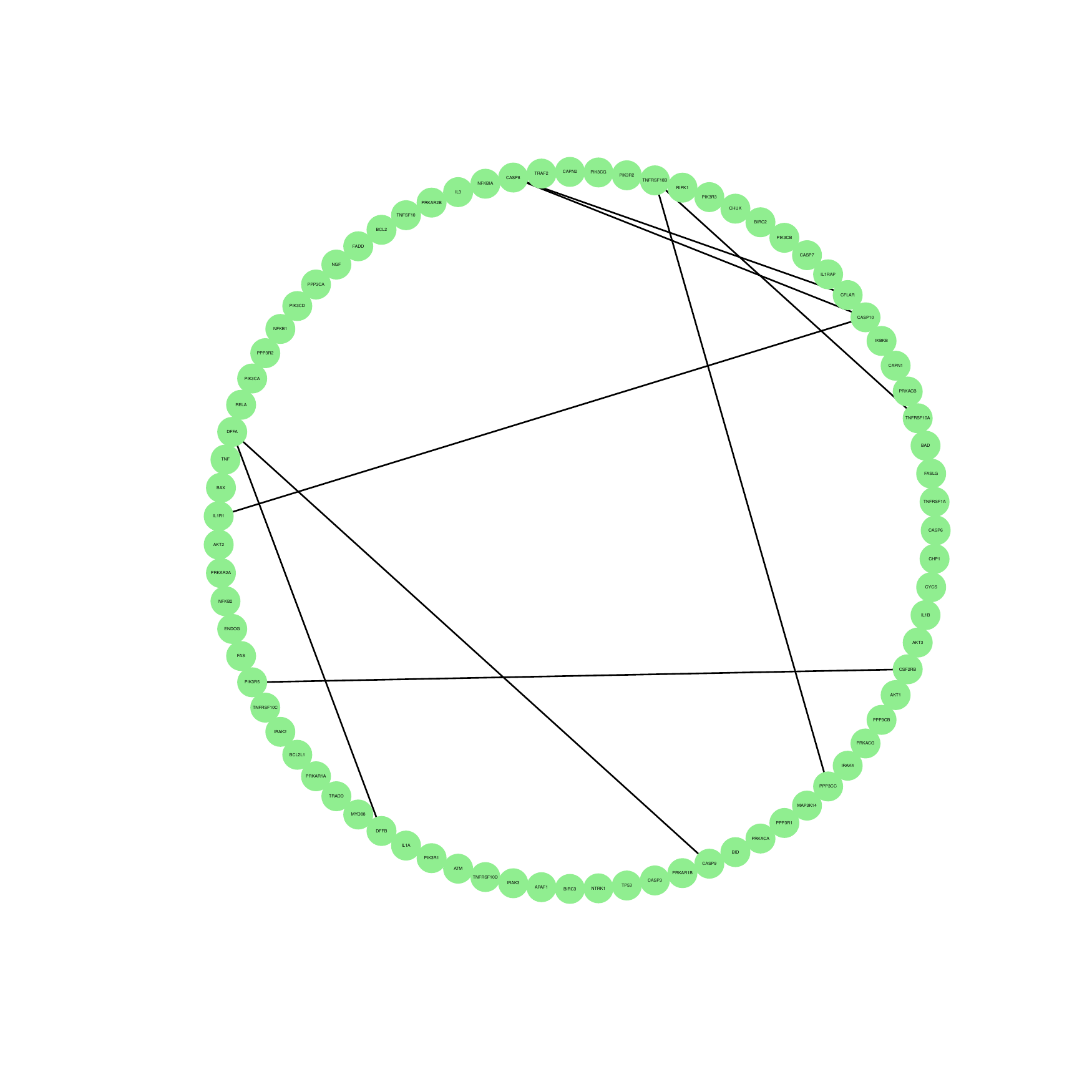}
  \end{minipage}\hfill
	\begin{minipage}[c]{0.33\linewidth}
    \centering
  			\includegraphics[trim=90 85 65 70,clip,width=1.00\textwidth]{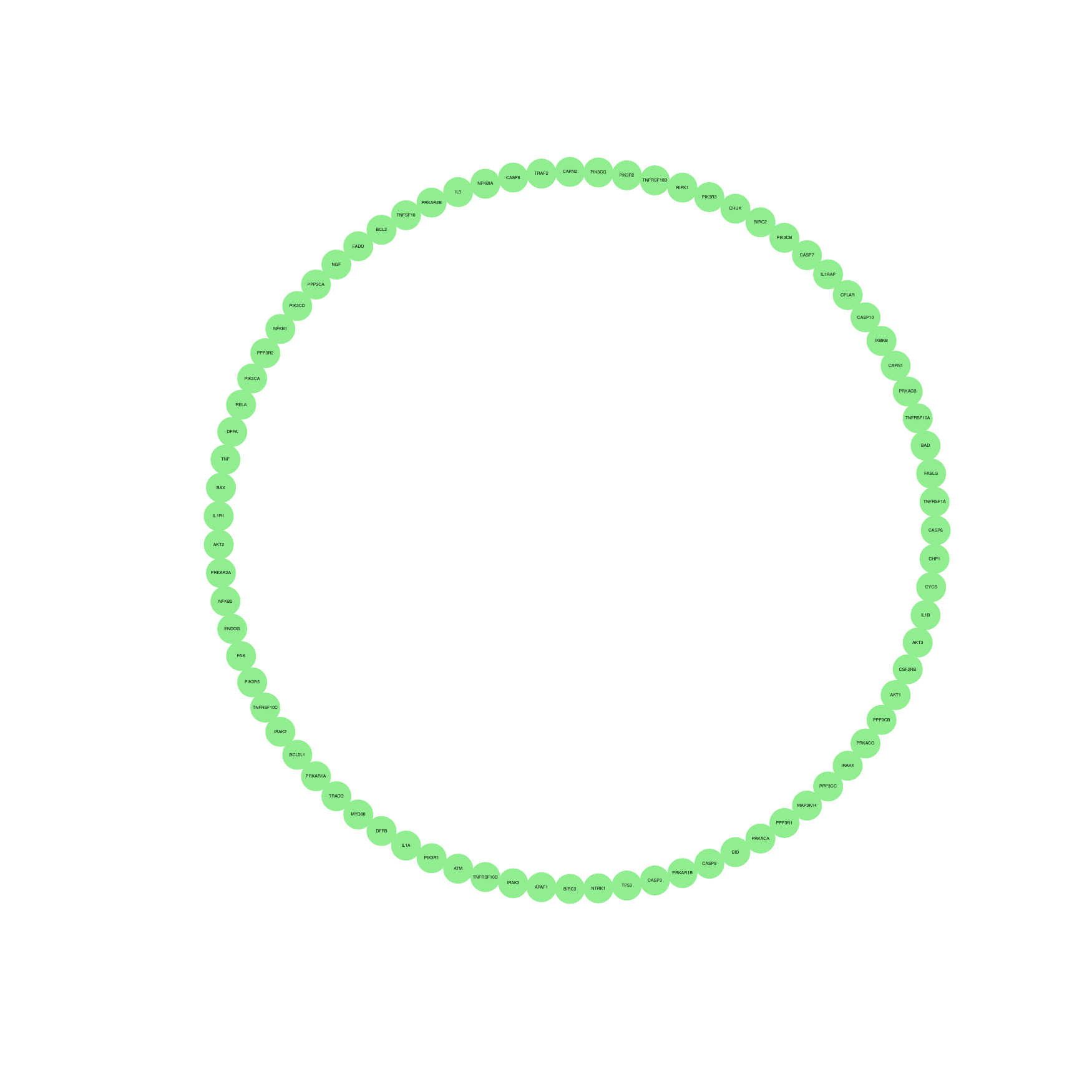}
  \end{minipage}\hfill
  \begin{minipage}[c]{0.33\linewidth}
    \centering
  			\includegraphics[trim=90 85 65 70,clip,width=1.00\textwidth]{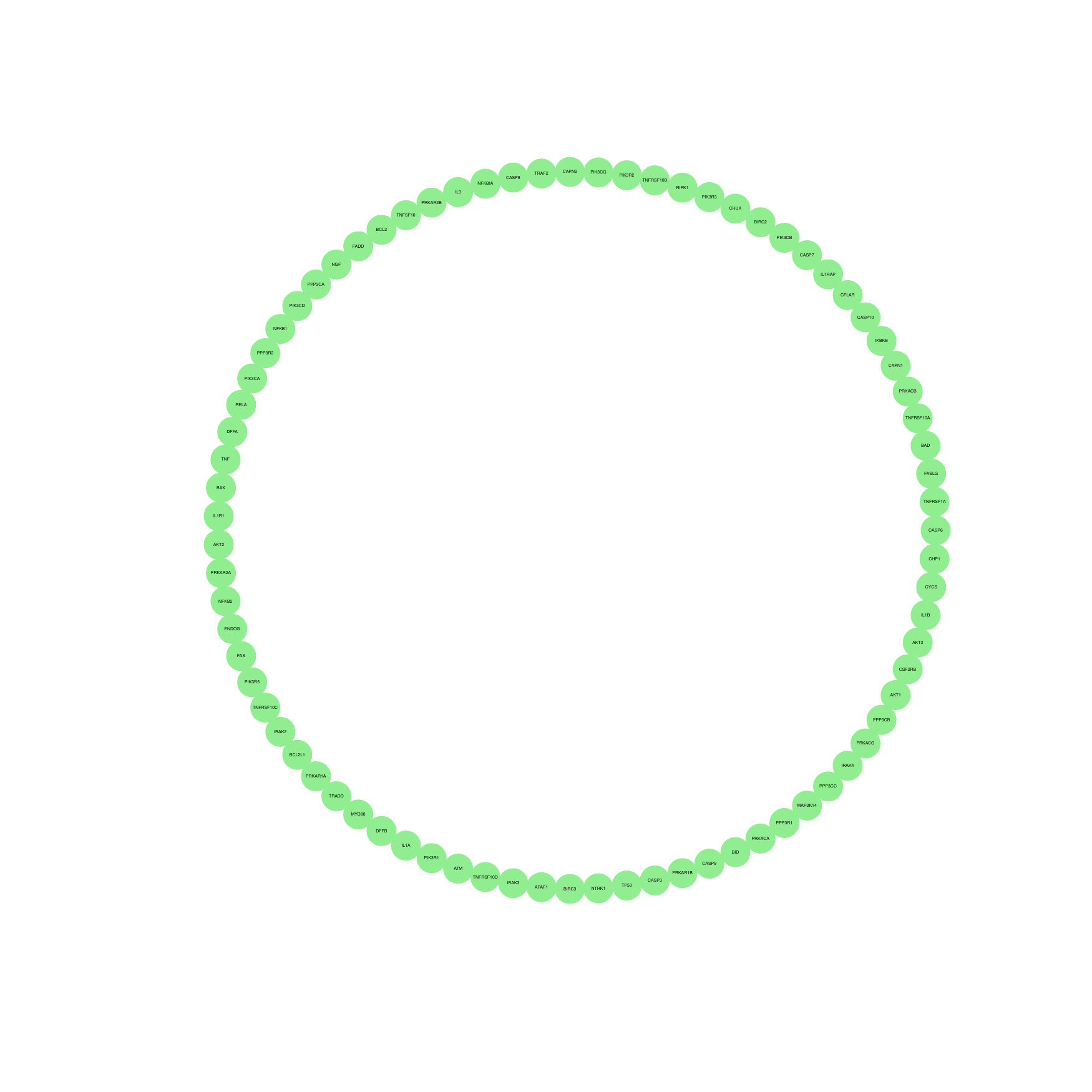}
  \end{minipage}\hfill
		\caption{Stable edges for the apoptosis pathway obtained with ShrinkNet (red), SEM$_\text{L}$ (blue), GL (pink) and GeneNet (green)  when $\mathbb{E}(V)=30$ as a function of $n_{\text{small}}^{\text{apopt}}$. Plots were generated using the R CRAN package \textit{rags2ridges}.}
		\label{Ugraphs:apopt}
\end{figure}

To illustrate, when $n_{\text{small}}^{\text{apopt}}=158$ for the apoptosis data we obtain that $\pi^{\text{ShrinkNet}}_{\text{thr}}=0.623$, $\pi^{\text{SEM}_\text{L}}_{\text{thr}}= 0.508$, $\pi^{\text{GL}_\lambda}_{\text{thr}}= 0.522$ and  $\pi^{\text{GeneNet}}_{\text{thr}}=0.503$, which result in 27, 12, 12 and 8 stables edges, respectively. These are illustrated in the left column of Figure \ref{Ugraphs:apopt}. As $\mathbb{E}(V)$ is fixed, the value of $\pi_{\text{thr}}$ only varies between methods because estimates of $q$ differ. This is intuitive: if the method selects a lot of (few) edges we expect $\pi_{\text{thr}}$ to be large (small).

Figure \ref{Ugraphs:apopt} and SM Figure 10 display stables edges obtained with each method as a function of $n_{\text{small}}^{\text{apopt}}$ and $n_{\text{small}}^{\text{p53}}$, respectively. For the two data sets ShrinkNet selects an important number of stable edges. This is particularly true for the apoptosis pathway where the method clearly yields more stable edges than SEM$_\text{L}$, GL$_{\lambda}$ and GeneNet in all situations. Specifically, when $n_{\text{small}}^{\text{apopt}}=79$ ShrinkNet identifies a nearly identical network to GL$_{\lambda}$ and SEM$_\text{L}$ when $n_{\text{small}}^{\text{apopt}}=158$. For the p53 pathway (see SM Figure 10), GL$_{\lambda}$ detects more stable edges than ShrinkNet when $n_{\text{small}}^{\text{p53}}=134$, as many as when $n_{\text{small}}^{\text{p53}}=67$, and less when $n_{\text{small}}^{\text{p53}}=40$. This suggests that when the sample size is small ShrinkNet tends to select more stable edges than GL$_{\lambda}$. Finally, for the two data sets ShrinkNet detects more stable edges than SEM$_\text{L}$ and GeneNet.

\newpage

\newpage

%%%%%%%%%%%%%%%%%%%%%%%%%%%%%%%%%%%%%%%%%%%%%%%%%%%%%%
%%%%%%%%%%%%%%%%%%%%%%%%%%%%%%%%%%%%%%%%%%%%%%%%%%%%%%
%%%	section: CONCLUSION
%%%%%%%%%%%%%%%%%%%%%%%%%%%%%%%%%%%%%%%%%%%%%%%%%%%%%%
%%%%%%%%%%%%%%%%%%%%%%%%%%%%%%%%%%%%%%%%%%%%%%%%%%%%%%

\section{Conclusion}
\label{conclusion}
In this paper we proposed a Bayesian SEM with global-local shrinkage priors for gene network reconstruction. The model employs simple conjugate priors to impose regularization. Because these are not sparse, a novel method for a posteriori edge selection was introduced to infer the graph structure. Computational efficiency was achieved by SVD decompositions and fast variational approximations. We discussed empirical Bayes estimation of prior hyper-parameters and embedded this in a variational EM-type algorithm. The simulations showed that the proposed approach is often superior to popular (sparse) methods in low-, moderate- and high-dimensional cases. In particular, on real data the method yielded more stable and reproducible discoveries.

A novelty of our work is the use of \emph{global} shrinkage priors, which allow the borrowing of information across regression equations. We are not aware of any previous works combining global and local shrinkage priors. In the frequentist setting \citet{yuan2012} borrows information across the regularizing parameters corresponding to $\ell_1$-penalties by combining local and global searches. In the Bayesian setting the focus is often on studying the equivalence between the SEM and a proper joint distribution \citep{geiger2002,dobra2004}. In this paper we have shown that the combined use of global and local shrinkage priors improves statistical inference, in particular edge ranking.

Our variable selection method performs simultaneous selection of the two parameters that are associated with each edge, but unlike sparsity-based methods performs separate estimation and selection steps. However, separating estimation and selection may also come as an advantage in terms of optimizing performance with respect to either of these criteria. In fact, ``The idea of pre-ranking covariates and then selecting models has become a well established technique in the literature'' \citep[Remark~6]{ishwaran2005}.

An important practical advantage of our approach is that the estimation procedure is coherent and complete, and does not rely on tuning, resampling, or cross-validation to set regularization parameter(s). This is particularly encouraging for extending the method to settings with multiple types of high-dimensional covariates, which would require different amounts of shrinkage. For methods based on resampling or cross-validation this may become overly computationally burdensome.

The proposed method is particularly suitable for gene network reconstruction using expression data. This type of network aims at providing a picture of regulatory mechanisms that act between genes.
In practice, the interest often lies in a relatively small subset of genes that are known to be functionally linked (e.g. a pathway).
In this context the Bayesian SEM may be more appropriate than others, because such a gene set is usually of moderate dimension and, hence, due to the functional link, the corresponding network is likely to be relatively less sparse. Therefore strong dependencies between genes are more likely to occur and this may favor Normal-Gamma (ridge-type) regularization. In addition, the coherence in functionality may render shrinkage beneficial for parameter estimation in the SEM.

We have focused on recovering the support of the precision matrix, but 
it is also possible to obtain an estimate of it. An immediate approach is to use the graph structure provided by the Bayesian SEM as a prior for precision estimation (sometimes referred to as  \emph{parameter learning} \citep{scutari2013}). Versions of the Wishart distribution, such as the G-Wishart \citep{dobra2011,wang2012}, are computationally attractive. Other estimation strategies have been proposed outside the Bayesian paradigm. These are usually based on the idea of thresholding. See, for example, \citet{zhou2011} and \citet{yuan2010}.

We foresee several extensions. SEMs are appropriate to describe directed networks and it would be interesting to investigate different types of shrinkage priors suitable in this context, for example to shrink in- and outgoing edges differently. Extension to non-Gaussian data is possible, where it may be desirable to adopt a flexible likelihood model and other types of posterior approximations may be considered \citep{rue2009}. Finally the model suits construction of integrative networks when allowing different priors for different types of interactions.\\

%\section*{Acknowledgements}

\begin{supplement}
%\sname{Supplement A}\label{suppA}
\stitle{Technical details and complementary results}
\slink[url]{DOI link here}
\sdescription{We present technical and methodological details regarding the variational approximation and the different methods under comparison in Sections \ref{modelSim} and \ref{dataSim}. Furthermore, complementary simulation results are provided.}
\end{supplement}

% BIBLIOGRAPHY
\bibliographystyle{imsart-nameyear}
\bibliography{references}

\end{document}